\documentclass[12pt, letterpaper]{article}
\usepackage[utf8]{inputenc}

\usepackage[top=1in, right = 1in, left = 1in, lines = 25]{geometry}

    \usepackage{paralist}
    \usepackage[compact]{titlesec}
   \titlespacing{\section}{0pt}{2ex}{1ex}
    \titlespacing{\subsection}{0pt}{1ex}{0ex}
    \titlespacing{\subsubsection}{0pt}{0.5ex}{0ex}

\usepackage[ruled,vlined]{algorithm2e}
\usepackage{mathtools}
\usepackage{graphicx}
\usepackage{multirow}
\usepackage{amssymb} 
\usepackage{amsthm}
\usepackage{amsmath}
\usepackage{fullpage}
\usepackage[longnamesfirst]{natbib}
\usepackage{enumerate}
\usepackage{amsfonts, bm, amsmath, color, natbib, rotating, amsthm, multicol, epstopdf, verbatim, hyperref,caption}  
\usepackage{authblk}
\usepackage{soul}
\usepackage{xr}
\bibpunct{(}{)}{;}{a}{,}{,} 

\usepackage{setspace}
\theoremstyle{plain} \newtheorem{theorem}{Theorem}  \newtheorem{lemma}{Lemma} \newtheorem{corollary}{Corollary}

\theoremstyle{definition}   

\theoremstyle{remark} \newtheorem*{remark}{Remark}  
\begin{document}

\title{
Subset selection for linear mixed models
}
\author{Daniel R. Kowal\thanks{
Dobelman Family Assistant Professor, Department of Statistics, Rice University, Houston, TX (\href{mailto:Daniel.Kowal@rice.edu}{daniel.kowal@rice.edu}).}}

\date{\today\vspace{-12mm}}

\maketitle
  

  \begin{abstract}
Linear mixed models (LMMs) are instrumental for regression analysis with structured dependence, such as grouped, clustered, or multilevel data. However, selection among the covariates---while accounting for this structured dependence---remains a challenge.  We introduce a Bayesian decision analysis for subset selection with LMMs. Using a Mahalanobis loss function that incorporates the structured dependence,   we derive optimal linear coefficients for (i) any given subset of variables and (ii) all subsets of variables that satisfy a cardinality constraint. Crucially, these estimates inherit shrinkage or regularization and uncertainty quantification from the underlying Bayesian model, and   apply for any well-specified Bayesian LMM. More broadly, our decision analysis strategy deemphasizes the role of a single ``best" subset, which is often unstable and limited in its information content, and instead favors a collection of near-optimal subsets. This collection is summarized by key member subsets and variable-specific importance metrics.  Customized subset search and out-of-sample approximation algorithms are provided for more scalable computing. These tools are applied to simulated data and a longitudinal physical activity dataset, and demonstrate excellent prediction, estimation, and selection ability. 
  \end{abstract}
  {\bf Keywords:} Bayesian analysis; hierarchical models; prediction; regression; variable selection

\thispagestyle{empty}

\clearpage
\setcounter{page}{1}

\section{Introduction}
Linear mixed models (LMMs) enable regression analysis in the presence of structured dependence, such as longitudinal data, grouped or clustered observations, or spatio-temporal effects. LMMs  are widespread in both Bayesian and classical statistical analysis and include many  hierarchical models and linear regression as special cases. We consider LMMs of the general form
\begin{equation}\label{lme}
\bm y = \bm X \bm \beta + \bm Z \bm u + \bm \epsilon,
\end{equation}
where $\bm y$ is the $N$-dimensional response, $\bm X$ is the $N \times p$ matrix of covariates, $\bm \beta$ is the $p$-dimensional vector of fixed effects regression coefficients, $\bm Z$ is the $N \times q$ random effects design matrix, $\bm u$ is the $q$-dimensional vector of random effects regression coefficients, and $\bm \epsilon$ is the $N$-dimensional observation error. Model \eqref{lme} is paired with the assumptions that  $\bm u$ and $\bm \epsilon$ are uncorrelated and mean zero with $\mbox{Cov}(\bm u) = \bm{\Sigma_u}$ and $\mbox{Cov}(\bm \epsilon) = \bm{\Sigma_\epsilon}$. Most commonly, the random effects $\bm u$ and the errors $\bm \epsilon$ are endowed with Gaussian distributions, but our approach does not require   any specific distributional assumptions beyond these moments.

The benefit of the LMM \eqref{lme} is that it marries the classical linear regression term $\bm X \bm \beta $ with a random effects term $\bm Z \bm u$ to capture structural dependence unexplained by $\bm X \bm \beta $. More formally, \eqref{lme} can be expressed in the marginal form 
$ 
\bm y = \bm X \bm \beta + \bm \nu, 
$  
where $\bm \nu \coloneqq \bm Z \bm u + \bm \epsilon$ has mean zero and covariance $\bm Z \bm{\Sigma_u} \bm Z' + \bm{\Sigma_\epsilon}$. The covariance of $\bm \nu$ incorporates elements of the random effects design $\bm Z$, the random effects covariance $\bm{\Sigma_u}$, and the observation error covariance $\bm{\Sigma_\epsilon}$.  LMMs are capable of modeling a broad variety of dependence structures; specific examples are given in Section~\ref{pda-lme}. 

Regardless of the structured dependence in the LMM, a core goal of regression analysis is \emph{selection} among the $p$ (fixed effects) covariates $\bm x$. Selection provides interpretable summaries of the data, reduced storage requirement, and often better prediction and lower estimation variability. We emphasize four main priorities that motivate our approach: 
\begin{enumerate}[(P1)]
\item The selection criteria and accompanying performance metrics should account for the \emph{structured dependence} modeled by the LMM; 
\item Selection should be applied \emph{jointly} across covariates rather than \emph{marginally} for each covariate;  
\item Selection of a \emph{single} ``best" subset of covariates should be  accompanied by an analysis of ``near-optimal" subsets of covariates; and 
\item The  inference and selection procedure should be computationally scalable in $N$ and $p$. 
\end{enumerate}
 P1 simply states that any structured dependence worth modeling in the LMM must also be included in the selection and evaluation process---which renders many existing tools ineligible.  P2 notes that variables selected using marginal criteria, such as hypothesis tests of the form $H_{0j}:\beta_j = 0$ or posterior inclusion probabilities from sparse Bayesian models, do not necessarily satisfy any joint optimality criteria. Hence, reporting the marginally-selected variables as a joint subset of variables often lacks justification.  More directly, P2 is satisfied only for subset selection. Yet subset selection is accompanied by other challenges, including selection instability and computational scalability.   P3 addresses the instability of subset selection: the ``best" subset often changes dramatically under minor perturbations or resampling of the data. This effect  is most pronounced in the presence of correlated covariates, weak signals, or small sample sizes, and undermines the elevated status of a ``best" subset. By instead collecting ``near-optimal" subsets, we acquire more information about the competing (predictive) explanations.  Lastly, P4 recognizes the computational burdens of subset search and demands tools that are feasible for moderate to large $N$ and $p$. 

Variable selection for LMMs has most commonly relied on penalized maximum likelihood estimation. \cite{Foster2007} and \cite{Wang2011} incorporated random effects within an adaptive lasso estimation procedure to account for genetic and experimental effects in quantitative trait loci  analysis and plant population studies, respectively. 
\cite{Bondell2010} and \cite{Ibrahim2011} selected fixed and random effects jointly using a modified Cholesky decomposition with adaptive lasso or SCAD penalties. These Cholesky parametrizations are order-dependent, so permutations of the columns of $\bm Z$ can produce different estimates and selections.  \cite{Muller2013} also noted that the accompanying algorithms can be slow and fail to converge, and reviewed alternative strategies such as information criteria. \cite{Fan2012} selected fixed effects by marginalizing over the random effects and maximizing a penalized (marginal) log-likelihood. The primary limitation is the need for a ``proxy matrix" for the inverse marginal covariance (of $\bm \nu$); \cite{Fan2012} simply used a multiple of the identity matrix, but this ignores the random effects covariance structure. In general, such penalized estimators can address priorities P1, P2, and P4, but not P3: they focus on selecting a single ``best" subset, and the accompanying (forward) search paths are too restrictive to enumerate a sufficiently rich collection of competitive subsets. 

From a Bayesian perspective, \cite{Chen2003} and \cite{Kinney2007} proposed sparsity-inducing spike-and-slab priors for both the fixed and random effects. These priors are compatible with our approach. The primary distinction is the mechanism for selection:  \cite{Chen2003} and \cite{Kinney2007}  compute posterior probabilities for all possible submodels. However, this strategy is computationally prohibitive and unreliable for small to moderate $p+q$, since only a small fraction of possible subsets can be visited regularly within the stochastic search Gibbs sampler. Hence, P4 is not satisfied. Marginal criteria such as posterior inclusion probabilities or hard-thresholding resolve these challenges, but fail to satisfy P2. 

 More broadly,  \cite{lindley1968choice} and \cite{hahn2015decoupling} have argued that selection is a \emph{decision problem} distinct from model specification. Sparsity or shrinkage priors cannot alone select variables: the prior is a component of the Bayesian model while the selection process requires its own criteria, typically a loss function that balances accuracy with sparsity. This decision analysis approach to selection has proven useful for functional regression \citep{kowal2019bayesianfosr}, 
seemingly unrelated regressions \citep{Puelz2017}, and
graphical models \citep{Bashir2019}, among others. However, these methods were not designed for LMMs and therefore fail to satisfy P1. In addition, with the exception of  \cite{Kowal2021}, these decision analysis approaches use (variations of) $\ell_1$-penalties and suffer from the same restrictive search paths as in classical penalized regression, which fails to satisfy P3. 

We propose a Bayesian approach for subset search and selection in LMMs that satisfies P1--P4. Using decision analysis  with a predictive loss function that directly incorporates the structured dependence in \eqref{lme}, we derive and compute the optimal linear coefficients for (i) any \emph{given} subset $\mathcal{S} \subseteq \{1,\ldots,p\}$ of variables and (ii)  all subsets of variables that satisfy a \emph{cardinality constraint} $\vert \mathcal{S} \vert \le k$ (P1, P2). These optimal coefficients are computable for any Bayesian LMM and inherit model-based regularization and posterior predictive uncertainty quantification. Linear coefficients are compared across subsets using out-of-sample predictive performance metrics that leverage both the structural dependencies and the predictive uncertainty from the Bayesian LMM. From these metrics, we construct the \emph{acceptable family} of near-optimal subsets, which collects those subsets that perform nearly as well as the ``best" subset with nonnegligble probability under the Bayesian LMM (P3). The acceptable family is more informative and robust than the ``best" subset---which itself is a member---and is summarized using other key member subsets and variable importance metrics. Customized subset search and out-of-sample approximation algorithms are provided to enable scalable computing (P4). 

We focus on subset selection of \emph{fixed effects} covariates, but note that the distinction between fixed and random effects is less pertinent for Bayesian modeling. Unlike frequentist LMMs that place a prior only on the random effects, Bayesian models require a prior on all parameters. Here, we consider ``fixed effects" as those covariates designated for selection, while ``random effects" capture the structured dependencies unmodeled by the fixed effects. 

The methodology is applied to moderate-to-vigorous physical activity (MVPA) data from the 2005-2006 National Health and Nutrition Examination Survey (NHANES). Repeated measurements of daily MVPA were recorded for each subject for one to seven days, along with several subject-specific demographic, health, and behavioral variables. The goal is to analyze which of these variables predict MVPA while adhering to priorities P1--P4 and accounting for the structured dependence implied by the longitudinal observations.

The paper is outlined as follows: Section~\ref{methods} develops the methodology and algorithms; Section~\ref{sims} provides results for simulated data; Section~\ref{app} presents an application to physical activity data; Section~\ref{concl} concludes. Supporting information includes a document with  additional simulation results, and additional results from the NHANES application, proofs of all results, and
computational details; and \texttt{R} code to reproduce the simulation study and data analysis. An \texttt{R} package is available at \url{https://github.com/drkowal/BayesSubsets}.





\section{Methods}\label{methods}

\subsection{Predictive decision analysis for linear mixed models}\label{pda-lme}
Bayesian analysis of LMMs pairs the model \eqref{lme} with suitable priors on $\bm \beta$ and $\bm u$ and a distributional choice for $\bm \epsilon$ to determine the likelihood, which is typically Gaussian. Specific choices will depend on the formulation of \eqref{lme} and are discussed subsequently; for now, we denote a generic Bayesian LMM by $\mathcal{M}$. The Bayesian model $\mathcal{M}$ induces a data-generating process via the posterior predictive distribution, 
\begin{equation}\label{predictive}
p_{\mathcal{M}}\{\bm{\tilde y}(\bm{\tilde X}, \bm{\tilde Z}) \mid \bm y\} = \int p_{\mathcal{M}}\{\bm{\tilde y}(\bm{\tilde X}, \bm{\tilde Z}) \mid \bm \theta\} \ p_{\mathcal{M}}(\bm \theta \mid \bm y)  \ d\bm\theta,
\end{equation}
 where $\bm \theta$ denotes the model $\mathcal{M}$ parameters including $\bm \beta$, $\bm u$, and any covariance parameters. The terms in the integrand are defined by the likelihood in \eqref{lme} evaluated at the covariate values $\bm{\tilde X}$ and $\bm{\tilde Z}$ and the joint posterior distribution under $\mathcal{M}$. Informally, \eqref{predictive} describes the distribution of future or unobserved data $\bm{\tilde y}$ at the design matrices $\bm{\tilde X}$ and $\bm{\tilde Z}$  conditional on the observed data $\bm y$ and according to model $\mathcal{M}$. The choice of $\bm{\tilde X}$ and $\bm{\tilde Z}$ can target covariate values or subpopulations of interest and determines the type of predictive observations, such as predictions for a new group or new measurements on an existing group. Absent other considerations, our default is  the observed matrices, $\bm{\tilde X} = \bm X$ and $\bm{\tilde Z} = \bm Z$

While the posterior predictive distribution formalizes the model-based uncertainty about unobserved data $\bm{\tilde y}(\bm{\tilde X}, \bm{\tilde Z})$, \emph{predictive decision analysis} determines the actions---point or interval predictions or estimators, selection among hypotheses, etc.---that provide optimal data-driven decision-making under $\mathcal{M}$.  Here, the goals are to (i) compute optimal linear coefficients for any given subset $\mathcal{S}\subseteq \{1,\ldots,p\}$ of variables, (ii) conduct an efficient search over candidates subsets,  and (iii) evaluate and compare predictive performance among subsets---all while adhering to the priorities P1--P4. Predictive decision analysis requires a loss function of the form $\mathcal{L}\{\bm{\tilde y}(\bm{\tilde X}, \bm{\tilde Z}), \bm \delta\}$, which enumerates the cost of an action $\bm \delta$ when $\bm{\tilde y}(\bm{\tilde X}, \bm{\tilde Z})$ is realized. In accordance with P1 and P2, we deploy a Mahalanobis loss function
\begin{equation}\label{loss}
\mathcal{L}\{\bm{\tilde y}(\bm{\tilde X}, \bm{\tilde Z}),  \bm \delta_\mathcal{S}; \bm\psi\} = \Vert \bm{\tilde y}(\bm{\tilde X}, \bm{\tilde Z}) - \bm{\tilde X} \bm \delta_\mathcal{S} \Vert_{\bm \Omega_{\bm \psi}}^2
\end{equation}
where $\bm \delta_\mathcal{S}$ is the $p$-dimensional  linear coefficients with zeros for any index $j \not\in\mathcal{S}$ and the norm $\Vert \bm v \Vert_{\bm \Omega_{\bm \psi}}^2 = \bm v' \bm \Omega_{\bm \psi} \bm v$ depends on a positive definite weighting matrix $\bm \Omega_{\bm \psi}$ that can depend on model parameters $\bm \psi$. 

For  LMMs, a natural choice of $\bm{\Omega_\psi}$ is the inverse marginal covariance of $\bm \nu$,
\begin{equation}\label{prec}
 \bm\Omega_{\bm \psi} = (\bm{\tilde Z} \bm{\Sigma_u}\bm{\tilde Z} ' + \bm{\Sigma_\epsilon})^{-1}
\end{equation}
with $\bm\psi = (\bm{\Sigma_u}, \bm{\Sigma_\epsilon})$. While the central quantity $\bm{\tilde y}(\bm{\tilde X}, \bm{\tilde Z}) - \bm{\tilde X} \bm \delta_\mathcal{S}$ in \eqref{loss} explicitly measures the linear predictive ability of a subset of variables  $\mathcal{S}$, the choice of \eqref{prec} incorporates weighting to account for the structured dependencies that are unknown yet modeled by the random effects under the LMM. With \eqref{prec}, the Mahalanobis loss \eqref{loss} resembles a multivariate Gaussian (negative) log-likelihood. However, this mathematical similarity should not be confused with a distributional assumption:  the Mahalanobis predictive loss \eqref{loss} inherits a joint posterior predictive distribution $p_{\mathcal{M}}\{\bm{\tilde y}(\bm{\tilde X}, \bm{\tilde Z}), \bm \psi \mid \bm y\}$ under $\mathcal{M}$.

For any \emph{given} subset $\mathcal{S}$, the optimal coefficients are obtained by minimizing the posterior expected loss under $\mathcal{M}$: 
\begin{equation}\label{action}
\bm{\hat \delta}_\mathcal{S} \coloneqq \arg\min_{\bm \delta_\mathcal{S}} \mathbb{E}_{[\bm{\tilde y}, \bm\psi \mid \bm y]} \mathcal{L}\{\bm{\tilde y}(\bm{\tilde X}, \bm{\tilde Z}), \bm \delta_\mathcal{S}; \bm \psi\},
\end{equation}
which averages over the joint uncertainty in  $\bm{\tilde y}(\bm{\tilde X}, \bm{\tilde Z})$ and $\bm \psi$ conditional on the data $\bm y$ and according to the model $\mathcal{M}$. The solution to \eqref{action} is derived explicitly: 
\begin{lemma}\label{thm-opt}
When $ \mathbb{E}_{[\bm{\tilde y}, \bm\psi \mid \bm y]} \Vert \bm{\tilde y}(\bm{\tilde X}, \bm{\tilde Z})\Vert_{\bm \Omega_{\bm\psi}}^2 < \infty$, the optimal coefficients in \eqref{action}  for a \emph{given} subset $\mathcal{S} \subseteq \{1,\ldots,p\}$ is given by the nonzero entries
\begin{equation}\label{action-solve}
\bm{\hat \delta}_\mathcal{S} =  (\bm {\tilde X}_\mathcal{S}'\bm{\hat \Omega} \bm{\tilde X}_\mathcal{S})^{-1} \bm{\tilde X}_\mathcal{S}'  \bm{\hat y^{\Omega}}
\end{equation}
with zeros for indices $j\not\in\mathcal{S}$, where $\bm{\tilde X}_\mathcal{S}$ subsets the columns of $\bm{\tilde X}$ based on $\mathcal{S}$ and $\bm{\hat \Omega} \coloneqq \mathbb{E}_{[\bm \psi \mid \bm y]} \bm\Omega_{\bm \psi}$ and $\bm{\hat y^{\Omega}} \coloneqq  \mathbb{E}_{[\bm{\tilde y}, \bm\psi \mid \bm y]}  \{\bm \Omega_{\bm \psi}\bm{\tilde y}(\bm{\tilde X}, \bm{\tilde Z})\}$ are posterior expectations under $\mathcal{M}$.
\end{lemma}
A generalized inverse may be substituted when  the solution \eqref{action-solve} is nonunique. 

Lemma~\ref{thm-opt} explicitly derives the optimal Bayesian estimator under Mahalanobis loss for any \emph{given} subset $\mathcal{S}$. The optimal  $\bm{\hat \delta}_\mathcal{S}$ is a ``fit to the fit" from $\mathcal{M}$, and therefore inherits shrinkage or  regularization from the Bayesian LMM. For illustration, consider a fixed and known weighting matrix $\bm \Omega$: the pseudo-response variable is $\bm{\hat y^{\Omega}} = \bm{\Omega}  \bm{\hat y}$  where $\bm{\hat y} \coloneqq  \mathbb{E}_{[\bm{\tilde y}  \mid \bm y]} \bm{\tilde y}(\bm{\tilde X}, \bm{\tilde Z})  = \bm{\tilde X} \bm{\hat \bm \beta} + \bm{\tilde Z} \bm{\hat u}$ for $\bm{\hat \beta} \coloneqq \mathbb{E}_{[\bm \beta \mid \bm y]} \bm \beta$ and $\bm{\hat u} \coloneqq \mathbb{E}_{[\bm u \mid \bm y]} \bm u$. The regularization from $\mathcal{M}$---usually applied via the priors for $\bm \beta$ and $\bm u$---is valuable for point prediction and estimation, and its absence in  classical subset selection is detrimental \citep{Hastie2020}.


The optimal coefficients in \eqref{action-solve} resemble generalized least squares (GLS) estimators for linear regression, including LMMs. The primary challenge in GLS estimation is that the inverse covariance or weight matrix $\bm{\Omega_\psi}$ is unknown. Feasible GLS iteratively estimates the covariance and the linear coefficients  via plug-in estimation, which is suboptimal. For LMMs, \cite{Fan2012} substituted a multiple of the identity matrix for $\bm{\Sigma_u}$ in \eqref{prec}  in order to avoid estimation of this covariance. These concessions are avoided in our approach: we solve a GLS optimization problem, but compute model-based expectations jointly over the unknown parameters---including the necessary inverse covariance matrix. The estimate of $\bm{\Omega_\psi}$ derives from the Bayesian LMM \eqref{lme}, which can benefit from the model-based regularization induced by the choice of shrinkage or sparsity priors under $\mathcal{M}$.

\subsection{The Mahalanobis weight matrix}\label{pda-weight}
To illustrate the use of the weighting matrix $\bm{\Omega_\psi}$, we consider several  examples. Since $\bm\Omega_{\bm \psi} = \bm{\Sigma_\epsilon}^{-1} - \bm{\Sigma_\epsilon}^{-1} \bm{\tilde Z} (\bm{\Sigma_u}^{-1} + \bm{\tilde Z} '\bm{\Sigma_\epsilon}^{-1}\bm{\tilde Z} )^{-1} \bm{\tilde Z} '\bm{\Sigma_\epsilon}^{-1}$ by the Woodbury identity, the common assumption of $\bm{\Sigma_\epsilon} = \sigma_\epsilon^2 \bm I_N$ results in the simplification
\begin{equation}
\label{maha-lme-2}
\bm\Omega_{\bm \psi}= \sigma_\epsilon^{-2} (\bm I_N - \bm{\tilde Z}  \bm{\Sigma_{u^*}}^{-1} \bm{\tilde Z} '),
\end{equation}
where $\bm{\Sigma_{u^*}} \coloneqq  \sigma_\epsilon^{2} \bm{\Sigma_u}^{-1} + \bm{\tilde Z} ' \bm{\tilde Z} $. The Mahalanobis predictive loss \eqref{loss} then decomposes as 
$$
\sigma_\epsilon^2\mathcal{L}\{\bm{\tilde y}(\bm{\tilde X}, \bm{\tilde Z}), \bm \delta_\mathcal{S}; \bm \psi\} = \Vert \bm{\tilde y}(\bm{\tilde X}, \bm{\tilde Z}) - \bm{\tilde X} \bm \delta_\mathcal{S} \Vert_2^2 - \Vert \bm{\tilde y}(\bm{\tilde X}, \bm{\tilde Z}) - \bm{\tilde X} \bm \delta_\mathcal{S} \Vert_{\bm{\tilde Z}  \bm{\Sigma_{u^*}}^{-1} \bm{\tilde Z} '}^2 
$$
which isolates the contribution from the squared error loss and the Mahalanobis loss based only on $\bm{\tilde Z} $ and $\bm{\Sigma_{u^*}}$---i.e., the critical terms in the random effects component.

The optimal coefficients in \eqref{action-solve} require computation of $\bm{\hat \Omega}$ and $\bm{\hat y^{\Omega}}$ under $\mathcal{M}$. We further consider two important examples: the random intercept model (Section~\ref{rand-int}) and the random slope model (Section~\ref{rand-slope}). 

\subsubsection{Random intercept model}\label{rand-int}
Consider longitudinal observations $\{y_{ij}\}_{j=1}^{m_i}$ on each subject $i=1,\ldots,n$, so $N = \sum_{i=1}^n  m_i$. The within-subject correlations are often modeled using the random intercept model
\begin{equation}\label{rand-int-model}
y_{ij} = \bm x_i' \bm \beta + u_i + \epsilon_{ij},
\end{equation}
usually with   $u_i \stackrel{iid}{\sim}N(0, \sigma_u^2)$  and $\epsilon_{ij} \stackrel{iid}{\sim} N(0, \sigma_\epsilon^2)$. The crucial role of $u_i$ cannot be ignored: since $\mbox{Corr}(y_{ij}, y_{ij'} \mid \bm x_i, \bm \beta) = \sigma_u^2/(\sigma_u^2 + \sigma_\epsilon^2)$, $\sigma_u$ accounts for the within-subject correlation that remains unexplained by the covariates $\bm x_i$. Model \eqref{rand-int-model} is a special case of \eqref{lme} with $\bm{\Sigma_u} = \sigma_u^2 \bm I_{N}$ and $\bm Z = \mbox{bdiag}\{\bm 1_{m_i}\}_{i=1}^n$  is a block diagonal matrix with $n$ $m_i$-dimensional vectors of ones. 

For predictive decision analysis, let $\bm{\tilde x}_i$ denote the target covariate values and $\tilde m_i$ the number of observations for each subject $i=1,\ldots, \tilde n$, which determines  $\bm{\tilde Z}$. The subject-specific predictive variables are 
$\bm{\tilde y}(\bm{\tilde X}, \bm{\tilde Z}) = (\bm{\tilde y}_1',\ldots,\bm{\tilde y}_{\tilde n}')'$ with $\bm{\tilde y}_i = (\tilde y_{i1},\ldots, \tilde y_{i\tilde m_i})'$ and the fixed effects covariate matrix is $\bm{\tilde X} = (\bm 1_{m_1}' \otimes \bm{\tilde x}_1, \ldots, \bm 1_{m_{\tilde n}}' \otimes \bm{\tilde x}_{\tilde n})'$. To compute $\bm{\Omega_\psi}$, observe that  $\bm{\Sigma_{u^*}} = \mbox{bdiag}\{\sigma_\epsilon^2/\sigma_u^2  + \bm 1_{\tilde m_i}'\bm 1_{\tilde m_i}\}_{i=1}^n = \mbox{diag}\{\sigma_\epsilon^2/\sigma_u^2  +\tilde m_i\}_{i=1}^n$ and  
$\bm{\Sigma_{u^*}}^{-1} = \mbox{diag}\{(\sigma_\epsilon^2/\sigma_u^2  +\tilde m_i)^{-1}\}_{i=1}^n$, so  the Mahalanobis weight matrix (up to $\sigma_\epsilon^2$) is 
\begin{equation}\label{prec-int}
\sigma_\epsilon^{2}\bm{\Omega_\psi} =  \mbox{bdiag}\Big\{ \bm I_{\tilde m_i} - \frac{1}{\sigma_\epsilon^2/\sigma_u^2  +\tilde m_i} \bm 1_{\tilde m_i} \bm 1_{\tilde m_i}' \Big\}_{i=1}^n
\end{equation}
 and does not require any numerical matrix inversions. Given \eqref{prec-int}, the Mahalanobis predictive loss simplifies to
\begin{equation}\label{maha-int}
\sigma_\epsilon^2 \Vert \bm{\tilde y} - \bm{\tilde X} \bm \delta_\mathcal{S} \Vert_{\bm \Omega_{\bm\psi}}^2 = \sum_{i=1}^{\tilde n} \Big[ \sum_{j=1}^{\tilde m_i} ( {\tilde y}_{ij} -  \bm{\tilde x}_i' \bm \delta_\mathcal{S})^2 -  \frac{1}{\sigma_\epsilon^2/\sigma_u^2  + \tilde m_i}  \Big\{\sum_{j=1}^{\tilde m_i} (\tilde y_{ij} - \bm{\tilde x}_i'\bm \delta_\mathcal{S})\Big\}^2\Big]
\end{equation}
which clearly isolates the difference between the Mahalanobis loss and squared error loss. In particular, \eqref{maha-int} incorporates the sign of the errors $e_{ij} \coloneqq {\tilde y}_{ij} -  \bm{\tilde x}_i' \bm \delta_\mathcal{S}$. For example, suppose $\tilde n=1$ and $\tilde m_1 = 2$, so the Mahalanobis loss (up to $\sigma_\epsilon^2$) is $e_{1}^2 + e_{2}^2 - (\sigma_\epsilon^2/\sigma_u^2 + 2)^{-1} (e_{1} + e_{2})^2$. The squared error loss $e_{1}^2 + e_{2}^2$ is invariant to the signs of the errors. However, the second term in \eqref{maha-int} includes a reduction in the loss by a factor of $(e_{1} + e_{2})^2$, which is larger when the errors have the same sign. Compared to the squared error loss, this Mahalanobis loss is more forgiving for errors in the same direction---and this is accentuated when $\sigma_u$ is large---which reflects the within-subject correlation induced by the underlying model \eqref{rand-int-model}.

  The posterior expectation $\bm{\hat \Omega}$ of \eqref{prec-int} is straightforward to compute, for example given posterior samples of $\{\sigma_\epsilon^2, \sigma_u^2\}$. To compute the posterior expectation of $\bm{\Omega_\psi}\bm{\tilde y}(\bm{\tilde X}, \bm{\tilde Z})$, the block diagonality simplifies this term to $n$ blocks of the form $\sigma_\epsilon^{-2} \{\bm I_{\tilde m_i} - (\sigma_\epsilon^2/\sigma_u^2  +\tilde m_i)^{-1} \bm 1_{\tilde m_i} \bm 1_{\tilde m_i}'\} \bm{\tilde y}_i = \sigma_\epsilon^{-2} \bm{\tilde y}_i - \{\sigma_\epsilon^{-2}(\sigma_\epsilon^2/\sigma_u^2  +\tilde m_i)^{-1}  \sum_{j=1}^{\tilde m_i} \tilde y_{ij} \}\bm 1_{\tilde m_i}.$ The posterior expectation of each $\tilde m_i$-dimensional vector is easily computable given posterior samples of $\{\sigma_\epsilon^2, \sigma_u^2, \bm{\tilde y}_i\}_{i=1}^{\tilde n}$.

\begin{remark}
These simplifications also provide a scalable Gibbs sampling algorithm for a Gaussian random intercept model with large $N,p$.  We apply a \emph{joint} sampling step for all fixed and random effects that (nearly) maintains the computational scalability of Bayesian linear regression \emph{without} the random intercepts. For simplicity, fix $m_i = m$  and let $\bm Y = [y_{ij}]$ denote the $n \times m$ matrix of observations. The strategy is to decompose  $[\bm\beta, \{u_i\}_{i=1}^n \mid \bm y, -] = [\bm\beta  \mid \bm y, -] [ \{u_i\}_{i=1}^n \mid \bm y, \bm\beta, -]$ and draw from the constituents of the product. Under the prior $\bm \beta \sim N(\bm 0, \bm \Sigma_\beta)$ and $u_i \stackrel{iid}{\sim}N(0, \sigma_u^2)$, the  regression coefficients satisfy $[\bm\beta  \mid \bm y, -] \sim N(\bm Q_\beta^{-1} \bm\ell_\beta, \bm Q_\beta^{-1})$ with $\bm Q_\beta = \omega^{tot}\bm X'\bm X + \bm \Sigma_\beta^{-1}$ and $\bm \ell_\beta = \bm X'\bm Y {\bm \omega_\cdot}^{tot}$, where $\omega^{tot} = \sum_{k,j} \bm \Omega_{kj}$ is the grand sum  and ${\bm \omega_\cdot}^{tot} = (\omega_{\cdot 1}^{tot},\ldots, \omega_{\cdot m}^{tot})'$ for $\omega_{\cdot j}^{tot} = \sum_{k=1}^m \bm \Omega_{kj}$  is the column sums of $\bm \Omega$ defined in \eqref{prec-int}. Notably, this distributional form matches the canonical posterior distribution of the regression coefficients for (non-LMM) Gaussian linear regression, which admits efficient sampling methods for large $n,p$ \citep{bhattacharya2016fast,Nishimura2018}. The random intercepts are sampled independently  via $[ u_i \mid \bm y, \bm\beta, -] \sim N(Q_{u_i}^{-1} \ell_{u_i}, Q_{u_i}^{-1})$ with $Q_{u_i} = m \sigma_\epsilon^{-2} + \sigma_u^{-2}$ and $\ell_{u_i} = \sigma_\epsilon^{-2} \sum_{j=1}^m (y_{ij} - \bm x_i'\bm\beta)$ for $i=1,\ldots,n$. Most important, these sampling steps for high-dimensional Bayesian random intercept regression are comparable to those for high-dimensional Bayesian linear regression, and only add minimal additional computations related to summations of $\bm \Omega$ and (parallelizable) draws of the scalar random intercepts $u_i$. The remaining sampling steps for the variance components are standard but depend on the choice of priors. These results also apply to Gibbs samplers for Gaussian mixture models (e.g., Dirichlet process mixtures of Gaussians) for $\epsilon_{ij}$ and/or $u_i$.  
\end{remark}

\subsubsection{Random slope model}\label{rand-slope}
Subject-specific slopes are common in hierarchical or multilevel models. By  applying \eqref{lme} with   $\bm Z = \mbox{bdiag}\{\bm x_i'\}_{i=1}^n$, the random slope model allows for subject-specific deviations from the population-level coefficients $\bm \beta$ (including a subject-specific intercept):
\begin{equation}\label{rand-slope-model}
y_i = \bm x_i' \bm \beta_i + \epsilon_i, \quad \bm \beta_i \coloneqq \bm \beta + \bm u_i.
\end{equation}
Model \eqref{rand-slope-model} is often accompanied by shrinkage priors on $\bm \beta$ and $\bm u_i$ to regularize against unnecessary predictors and unnecessary heterogeneity, respectively. Predictive decision analysis with Mahalanobis loss enables coefficient estimation and subset selection for $\bm x$ (see Section~\ref{sec-accept}) while adjusting for the heterogeneities induced by the random effects $\bm u_i$.  
 
When  $\bm{\Sigma_\epsilon} = \sigma_\epsilon^2 \bm I_N$, the key term $\bm{\Sigma_{u^*}}^{-1}$ in the inverse covariance \eqref{maha-lme-2} is directly available from the Sherman-Morrison formula, $\bm{\Sigma_{u^*}}^{-1} = \sigma_\epsilon^{-2} 
\mbox{bdiag}\big\{
\bm{\Sigma_{u_i}} - \bm{\Sigma_{u_i}} \bm{\tilde x}_i \bm{\tilde x}_i'\bm{\Sigma_{u_i}}/ (\sigma_\epsilon^2 + \bm{\tilde x}_i'\bm{\Sigma_{u_i} \bm{\tilde x}_i})
\big\}_{i=1}^n$.  The accompanying Mahalanobis weight matrix then simplifies to the diagonal matrix
$
\bm{\Omega_\psi} = \mbox{diag}\{\omega_i\}_{i=1}^n
$ with $\omega_i \coloneqq 1/(\sigma_\epsilon^2  + \bm{\tilde x}_i'\bm{\Sigma_{u_i}} \bm{\tilde x}_i)$,  which is computable without numerical matrix inversions. The implied Mahalanobis predictive loss is the weighted least squares $
\Vert \bm{\tilde y}(\bm{\tilde X}, \bm{\tilde Z}) - \bm{\tilde X} \bm \delta_\mathcal{S} \Vert_{\bm \Omega_{\bm \psi}}^2 =  \sum_{i=1}^{\tilde n} \omega_i \{\tilde y_i(\bm{\tilde x}_i) - \bm{\tilde x}_i' \bm \delta_\mathcal{S}\}^2$. The subject-specific weights $\omega_i$ are primarily driven by $ \bm{\tilde x}_i'\bm{\Sigma_{u_i}} \bm{\tilde x}_i$, where $\bm{\Sigma_{u_i}}$ is the covariance of the subject-specific deviations $\bm u_i = \bm \beta_i - \bm \beta$.  The posterior expectations required by Lemma~\ref{thm-opt} are straightforward: $\bm{\Omega_\psi} = \mbox{diag}\{\omega_i\}_{i=1}^n$ and $\bm{\Omega_\psi}\bm{\tilde y}(\bm{\tilde X}, \bm{\tilde Z})$ is an $\tilde n$-dimensional vector with elements $\{\omega_i \tilde y_i(\bm{\tilde x}_i)\}_{i=1}^{\tilde n}$, both of which are easily computable given posterior samples of $\{\sigma_\epsilon^2, \bm{\Sigma_{u_i}}, \tilde y_i(\bm{\tilde x}_i)\}_{i=1}^{\tilde n}$. 


\subsection{Subset search for linear mixed models}\label{pda-search}
Although Lemma~\ref{thm-opt} produces the optimal linear coefficients for a \emph{given} subset $\mathcal{S}$, it does not guide the subset \emph{search} or \emph{selection} process. To remedy this, we append the Mahalanobis loss function \eqref{loss} with a \emph{cardinality constraint} and define the optimal action
\begin{equation}
\label{sq-loss-action-card-1}
\bm{\hat \delta}_k \coloneqq \arg\min_{\bm \delta} \mathbb{E}_{[\bm{\tilde y}, \bm\psi \mid \bm y]} \mathcal{L}\{\bm{\tilde y}(\bm{\tilde X}, \bm{\tilde Z}), \bm \delta; \bm \psi\} \quad \mbox{subject to} \quad \Vert \bm \delta \Vert_0 \le k
\end{equation}
so $\bm{\hat \delta}_k$ provides the optimal coefficients among all subsets with \emph{at most} $k$ variables. The solution in \eqref{sq-loss-action-card-1} resemble the ``best subset selection" problem in classical regression (e.g., \citealp{Miller1984}), suitably modified for Bayesian decision analysis. 

The cardinality constraint diverges from the ubiquitous strategy among decision analysis methods for variable selection, which is  to append  the loss function 
(e.g., \eqref{loss}) with an $\ell_1$-penalty to encourage sparsity among the coefficients \citep{hahn2015decoupling}. Such a strategy may be viewed as a convex relaxation of \eqref{sq-loss-action-card-1}. 
However, the  $\ell_1$-penalty introduces additional regularization---beyond the regularization from $\mathcal{M}$---and can overshrink true signals. Adaptive lasso-type adjustments are available \citep{KowalPRIME2020} but cannot circumvent this issue entirely. Further, the (adaptive) lasso-based search paths are highly constrained within the space of all possible subsets, and therefore cannot enumerate a sufficiently broad collection of competitive subsets to satisfy P3.

We instead target \eqref{sq-loss-action-card-1} directly, and provide a substantial simplification of the solution:
 \begin{theorem}\label{sq-loss-action-card}
The optimal coefficients \eqref{sq-loss-action-card-1} using the loss \eqref{loss}  and the cardinality constraint   $\Vert \bm \delta \Vert_0 \le k$ (with $k \le p$) are 
\begin{equation}
 \label{sq-loss-action-card-2}
 \bm{\hat \delta}_k=  \arg\min_{\bm \delta} \Vert \bm{y}^* - \bm X^* \bm \delta\Vert_2^2    \quad \mbox{subject to} \quad \Vert \bm \delta \Vert_0 = k
\end{equation}
where $\bm{y}^* \coloneqq \bm{\hat \Omega}^{-1/2} \bm{\hat y^{\Omega}}$,  $\bm X^* \coloneqq \bm{\hat \Omega}^{1/2} \bm{\tilde X}$, and $(\bm{\hat \Omega}^{1/2})'\bm{\hat \Omega}^{1/2} = \bm{\hat \Omega}$. 
\end{theorem}
The expected predictive Mahalanobis loss in \eqref{sq-loss-action-card-1} is reduced to a squared error loss involving  pseudo-data $\bm{y}^*$ and $\bm X^*$. Most important, the squared error representation in \eqref{sq-loss-action-card-2} enables application of state-of-the-art subset search algorithms for \emph{classical} linear regression \citep{Furnival2000,Bertsimas2016} to the setting of  \eqref{sq-loss-action-card-1}. The pseudo-data $\bm{y}^*$ and $\bm X^*$ are a one-time computing cost, while the matrix square root $\bm{\hat \Omega}^{1/2}$ often admits fast Cholesky decompositions (e.g., block diagonality in Section~\ref{rand-int}) or direct computations (e.g., diagonality in Section~\ref{rand-slope}) depending on the form of the LMM \eqref{lme}. In addition, Theorem~\ref{sq-loss-action-card} reduces the search space from $2^k$ subsets to $p\choose{k}$  subsets. For any subset of size $k$, we simply apply Lemma~\ref{thm-opt} to compute the optimal linear coefficients as in \eqref{action-solve}.

Despite these advantageous results, Theorem~\ref{sq-loss-action-card} also highlights the limitations of the representation in  \eqref{sq-loss-action-card-1}. 
First, this solution does not consider P3: there may be many near-optimal subsets of smaller sizes, yet all subsets with $\vert \mathcal{S}\vert < k$  are immediately discarded. Second, this solution does not favor parsimony: the optimal coefficients $ \bm{\hat \delta}_k$ are the \emph{largest} allowable subset under the cardinality constraint. 
Hence, optimizing over all possible subsets is achieved by setting $k=p$, which yields a trivial solution: 
\begin{corollary}\label{opt-subset}
The optimal coefficients   under the loss \eqref{loss}  and computed \emph{across all possible subsets} $\mathcal{S} \subseteq \{1,\ldots,p\}$ are 
$\bm{\hat \delta}_{\mathcal{\widehat S}} \coloneqq  \arg\min_{\mathcal{S}, \bm \delta} \mathbb{E}_{[\bm{\tilde y}, \bm\psi \mid \bm y]} \mathcal{L}\{\bm{\tilde y}(\bm{\tilde X}, \bm{\tilde Z}), \bm \delta; \bm \psi\} = \bm{\hat \delta}_{\{1,\ldots,p\}} =  (\bm {\tilde X}'\bm{\hat \Omega} \bm{\tilde X})^{-1} \bm{\tilde X}'  \bm{\hat y^{\Omega}}$
with   $\mathcal{\widehat S} = \{1,\ldots,p\}$. 
\end{corollary}
Clearly, selection via direct optimization is inadvisable: the selected subset includes all variables and therefore is invariant to the data or the model. 

In conjunction,  Theorem~\ref{sq-loss-action-card} and Corollary~\ref{opt-subset} imply the need to (i) search over multiple cardinalities   $k=1,\ldots,p$ and (ii) develop alternative metrics to compare subsets of distinct sizes. Even with the simplifications of Theorem~\ref{sq-loss-action-card} and the accompanying   subset search algorithms  \citep{Furnival2000,Bertsimas2016}, it is often necessary to restrict the search space when $p$ is moderate or large. We do so by bounding the maximum subset size $s_{max} \le p$ and  the number of subsets $s_k \le {p\choose k}$ of each size $k$. 

First, we pre-screen to select the $s_{max} = \min\{p, 35\}$ covariates that have the largest effect sizes under the LMM. Although this pre-screening applies a \emph{marginal} criterion, it is based on a \emph{joint} model under $\mathcal{M}$.  In that sense, this procedure is similar to the most popular Bayesian variable selection strategies based on posterior inclusion probabilities  or hard-thresholding. In our case, this is a coarse pre-screening technique, not a terminal selection procedure.

Second, we apply the \emph{branch-and-bound algorithm} (BBA; \citealp{Furnival2000}) to filter to the ``best" $s_k$ subsets of each size $k=1,\ldots, s_{max}$. BBA searches through a tree-based enumeration of all possible subsets (up to size $s_{max}$),  yet avoids an exhaustive subset search by carefully eliminating non-competitive subsets (or branches) according to least squares. Hence, application of BBA requires a least squares representation of the expected predictive Mahalanobis loss in \eqref{sq-loss-action-card-1}, which we provide below:
\begin{lemma}\label{rss-ord-lem}
Let $\bm \delta_1$ and $\bm \delta_2$ denote linear coefficients. When $ \mathbb{E}_{[\bm{\tilde y}, \bm\psi \mid \bm y]} \Vert \bm{\tilde y}(\bm{\tilde X}, \bm{\tilde Z})\Vert_{\bm \Omega_{\bm\psi}}^2 < \infty$, we have the ordering equivalence
$ 
\mathbb{E}_{[\bm{\tilde y}, \bm\psi \mid \bm y]} \mathcal{L}\{\bm{\tilde y}(\bm{\tilde X}, \bm{\tilde Z}), \bm \delta_1; \bm \psi\} \le \mathbb{E}_{[\bm{\tilde y}, \bm\psi \mid \bm y]} \mathcal{L}\{\bm{\tilde y}(\bm{\tilde X}, \bm{\tilde Z}), \bm \delta_2; \bm \psi\}
$ 
if and only if
$ 
\Vert \bm{y}^* - \bm X^* \bm \delta_1\Vert_2^2 \le \Vert \bm{y}^* - \bm X^* \bm \delta_2 \Vert_2^2.
$ 
\end{lemma}
The key implication of Lemma~\ref{rss-ord-lem} is that we may directly apply BBA  using the pseudo-data $\bm{y}^*$ and $\bm X^*$ (defined in Theorem~\ref{sq-loss-action-card}) to obtain the ``best" $s_k$ subsets of each size $k=1,\ldots,s_{max}$. Ideally, $s_k$ should be set to the largest size possible for  a given computing budget. We use the default values   $s_k = 15$ or $s_k = 100$ and apply the efficient BBA implementation  in the  \texttt{leaps} package in \texttt{R}. However, Lemma~\ref{rss-ord-lem} also enables any other subset search strategy based on least squares (e.g., \citealp{Bertsimas2016}).

\subsection{Acceptable families for near-optimal subsets}\label{sec-accept}
Subset selection via the decision analysis in \eqref{sq-loss-action-card-1} is incomplete: the solution returns only the ``best" model of each size $k$ and trivially prefers the largest possible subset. Additional tools are needed to (i) compare subsets of distinct sizes and (ii) collect the \emph{near-optimal} subsets in accordance with P3.
For these tasks, we use \emph{out-of-sample} predictive performance and  adapt the \emph{acceptable family} of \cite{Kowal2020target} for the LMM setting. 
Informally, the acceptable family is the collection of all subsets that (nearly) match  the  predictive performance of the ``best" subset with nonnegligble probability under $\mathcal{M}$.  By studying this collection of near-optimal subsets, we deemphasize the role of a single ``best" subset in favor of many distinct yet predictively-competitive alternatives. The acceptable family has been applied for Bayesian subset selection 
 \citep{Kowal2021},  $\ell_1$-penalized selection \citep{KowalPRIME2020}, and targeted variable selection \citep{Kowal2020target}, but none have considered LMMs.
 
The acceptable family is built by evaluating out-of-sample predictive performance, which requires careful consideration for LMMs.  For repeated or longitudinal observations, it must be determined whether to evaluate predictions for new subjects or for new measurements on existing subjects. For concreteness, we proceed under the longitudinal setting of Section~\ref{rand-int} and evaluate predictions on new subjects. Modifications for other cases are available. 

Consider  $n$ subjects with $m_i$ observations per subject, $i=1,\ldots, n$.  We implement a Bayesian $K$-fold cross-validation procedure, where the $K$ folds are taken across subjects $i=1,\ldots,n$. Let $\mathcal{I}_k \subset \{1,\ldots,n\}$ denote the $k$th validation set, where each subject point appears in one validation set, $\cup_{k=1}^K \mathcal{I}_k = \{1,\ldots,n\}$. By default, we use $K=10$ validation sets that are equally-sized, mutually exclusive, and selected randomly from $\{1,\ldots,n\}$. 
For each subset $\mathcal{S}$, we define the out-of-sample \emph{empirical loss}
\begin{equation}
\label{emp-loss}
\mathcal{L}_\mathcal{S} \coloneqq \frac{1}{K} \sum_{k=1}^K \mathcal{L}_{\mathcal{S}}(k), \quad 
\mathcal{L}_{\mathcal{S}}(k) \coloneqq  \frac{1}{\vert \mathcal{I}_k \vert} \mathcal{L}( \bm y_{\mathcal{I}_k}, \bm{\hat \delta}_{\mathcal{S}}^{-\mathcal{I}_k}; \bm{\hat \psi}^{-\mathcal{I}_k}),
\end{equation}
where $\bm y_{\mathcal{I}_k} \coloneqq \{\bm y_i\}_{i \in \mathcal{I}_k}$ denotes the response variables on the validation data with $\bm y_i = (y_{i1},\ldots,y_{im_i})'$, 
$\bm{\hat \delta}_{\mathcal{S}}^{-\mathcal{I}_k} \coloneqq \arg\min_{\bm \delta_\mathcal{S}} \mathbb{E}_{[\bm{\tilde y}, \bm\psi \mid \bm y_{-\mathcal{I}_k}]} \mathcal{L}(\bm{\tilde y}_{\mathcal{I}_k}, \bm \delta_\mathcal{S}; \bm \psi\}$ are the optimal coefficients \eqref{action} but 
estimated using only the training data  $\bm y_{-\mathcal{I}_k} \coloneqq \{ \bm y_i\}_{i \not \in \mathcal{I}_k}$, and, with abuse of notation, $\bm{\hat \psi}^{-\mathcal{I}_k}$ in \eqref{emp-loss} indicates the Mahalanobis loss \eqref{loss} with weighting matrix $\bm{\hat\Omega}^{-\mathcal{I}_k} \coloneqq \mathbb{E}_{[\bm \psi \mid \bm y_{-\mathcal{I}_k}]} \bm\Omega_{\bm \psi}$. The empirical loss \eqref{emp-loss} resembles classical $K$-fold cross-validation with a point estimate for each Mahalanobis loss weighting matrix.  From this quantity, we define the ``best" subset for out-of-sample point prediction, 
 \begin{equation}\label{best}
\mathcal{S}_{min} \coloneqq \arg\min_{\mathcal{S} } \mathcal{L}_\mathcal{S},
\end{equation}
so that $\bm{\hat \delta}_{\mathcal{S}_{min}}$ are the optimal linear coefficients for the subset $\mathcal{S}_{min}$ that minimizes \eqref{emp-loss}.  

 To define the acceptable family, we first introduce the out-of-sample \emph{predictive loss} analogous to \eqref{emp-loss}:
\begin{equation}
\label{pred-loss}
\widetilde{\mathcal{L}}_{\mathcal{S}} \coloneqq \frac{1}{K} \sum_{k=1}^K \widetilde{\mathcal{L}}_{\mathcal{S}}(k), \quad 
\widetilde{\mathcal{L}}_{\mathcal{S}}(k) \coloneqq  \frac{1}{\vert \mathcal{I}_k \vert} \mathcal{L}( \bm{\tilde y}_{\mathcal{I}_k}^{-\mathcal{I}_k}, \bm{\hat \delta}_{\mathcal{S}}^{-\mathcal{I}_k}; \bm{\psi}^{-\mathcal{I}_k})
\end{equation}
where $ \bm{\tilde y}_{\mathcal{I}_k}^{-\mathcal{I}_k} \sim p_\mathcal{M}[ \{\bm{\tilde y}(\bm x_i, \bm z_i)\}_{i \in \mathcal{I}_k} \mid \bm y_{-\mathcal{I}_k}]$ denotes the predictive variables in  the validation set conditional on the training data and $\bm{\psi}^{-\mathcal{I}_k}$ similarly conditions only on the training data. Unlike the empirical loss ${\mathcal{L}}_{\mathcal{S}}$, the predictive loss  $\widetilde{\mathcal{L}}_{\mathcal{S}}$ incorporates out-of-sample predictive \emph{uncertainty} under $\mathcal{M}$, as well as the uncertainty regarding relevant model parameters $\bm \psi$. The uncertainty reflects the fact that the validation data $\bm y_{\mathcal{I}_k}$ represent only one possible realization of observables at the covariate values $\{\bm x_i, \bm z_i\}_{i \in \mathcal{I}_k}$. The (out-of-sample) predictive distribution provides alternative model-based realizations, and hence is informative for quantifying the uncertainty of out-of-sample predictive performance. 

Using the predictive loss, the acceptable family is defined as those subsets that are ``near-optimal" relative to $\mathcal{S}_{min}$:
 \begin{equation}\label{accept}
\mathbb{A}_{\eta, \varepsilon} \coloneqq \big\{ \mathcal{S}: \mathbb{P}_\mathcal{M}\big(\widetilde{\mathcal{D}}_{\mathcal{S}_{min},\mathcal{S}}  < \eta \big) \ge \varepsilon \big\}, \quad \eta \ge 0, \varepsilon \in [0,1]
\end{equation}
where $\widetilde{\mathcal{D}}_{\mathcal{S}_{min},\mathcal{S}}  \coloneqq 100\times(\widetilde{\mathcal{L}}_{\mathcal{S}} - \widetilde{\mathcal{L}}_{\mathcal{S}_{min}})/\widetilde{\mathcal{L}}_{\mathcal{S}_{min}}$ is the percent increase in predictive loss for subset $\mathcal{S}$ relative to $\mathcal{S}_{min}$, $\eta \ge 0\%$ is the margin, and $\varepsilon \in [0,1]$ is the probability level. Equivalently, a subset $\mathcal{S}$ is acceptable  if and only if there exists a lower $(1- \varepsilon)$ posterior prediction interval for $\widetilde{\mathcal{D}}_{\mathcal{S}_{min},\mathcal{S}}$ that includes $\eta$ \citep{Kowal2020target}. Subsets are \emph{not} acceptable if there is insufficient predictive probability under $\mathcal{M}$ that the out-of-sample accuracy is within a predetermined margin of the ``best" subset. $\mathcal{S}_{min}$ is necessarily a member of $\mathbb{A}_{\eta, \epsilon}$  for any $(\eta, \varepsilon)$, so the  acceptable family is always nonempty. Larger values of $\eta$ and smaller values of $\varepsilon$ expand the acceptable family; we select  $\eta = 0$ and $\varepsilon = 0.10$ by default  and conduct sensitivity analyses (see also \citealp{Kowal2020target,KowalPRIME2020,Kowal2021} for further sensitivity evaluations).

The acceptable family is related to fence methods for model selection \citep{Jiang2008}, which seek to eliminate ``incorrect" models using likelihood criteria. These methods rely on asymptotic arguments or bootstrap computations, while our approach emphasizes out-of-sample predictive performance with (finite-sample) predictive uncertainty quantification under the LMM. Fence methods primarily focus on selection of a \emph{single} model, rather than analysis of the \emph{collection} of near-optimal models or subsets. Most critically, in our empirical examples the fence methods failed to converge for $p \ge 10$ (using the \texttt{R} package \texttt{fence}), while the proposed approach is highly scalable in both $n$ and $p$.


We summarize the acceptable family using two strategies. First, we report two key members: the ``best" subset $\mathcal{S}_{min}$ and the \emph{smallest} acceptable subset, 
\begin{equation}\label{min-accept}
\mathcal{S}_{small} \coloneqq \arg\min_{\mathcal{S} \in \mathbb{A}_{\eta, \varepsilon}} \vert \mathcal{S}\vert,
\end{equation}
which is  the smallest subset of covariates that satisfies the near-optimality condition in \eqref{accept}. 
Typically, we find $\vert \mathcal{S}_{small}\vert \ll \vert \mathcal{S}_{min}\vert$ which is expected: selection based on minimizing cross-validation error is known to produce models that are unnecessarily complex. 
Although we caution against overreliance on any single subset, $\mathcal{S}_{small}$ is a uniquely valuable summary of the acceptable family: smaller subsets are \emph{not} acceptable, and therefore $\mathcal{S}_{small}$ offers a notion of the ``necessary" variables for near-optimal prediction. When $\mathcal{S}_{small}$  is nonunique, the acceptable subsets of size $\vert \mathcal{S}_{small}\vert$ should be evaluated in concert; a unique choice of $\mathcal{S}_{small}$ is given by the acceptable subset of this size that achieves the 
smallest empirical loss \eqref{emp-loss}.


Second, we summarize $\mathbb{A}_{\eta,\varepsilon}$ using the variable importance metric for each covariate $j$: 
\begin{equation}\label{vi-co}
\mbox{VI}_{\rm incl}(j)  \coloneqq  \vert \mathbb{A}_{\eta, \varepsilon}\vert^{-1} \sum_{\mathcal{S} \in \mathbb{A}_{\eta, \varepsilon}} \mathbb{I}\{j \in \mathcal{S}\},
\end{equation}
which can also be generalized for two or more covariates  \citep{Kowal2021}. This quantity is most informative at each endpoint: $\mbox{VI}_{\rm incl}(j) \approx 1$ implies that covariate $j$ belongs to (nearly) all acceptable subsets and is therefore an essential or \emph{keystone} covariate, while $\mbox{VI}_{\rm incl}(j) \approx 0 $ suggests that covariate $j$ is irrelevant for  (nearly) all acceptable subsets.   By design,  $\mbox{VI}_{\rm incl}(j) $ provides a \emph{variable-specific} summary of the acceptable family of subsets.  This metric is broadly related to stability selection
\citep{meinshausen2010stability} and frequentist aggregation of variable importance across many ``good" models \citep{Dong2019}.


To compute the out-of-sample quantities in \eqref{emp-loss} and \eqref{pred-loss} under $\mathcal{M}$, we use an importance sampling algorithm. This algorithm requires only the \emph{in-sample} posterior under the LMM and hence avoids the intensive processing of re-fitting $\mathcal{M}$ for each of the $K$ folds. The algorithm is detailed in the supporting information and modifies previous approaches \citep{Kowal2020target,KowalPRIME2020,Kowal2021} for LMMs and Mahalanobis loss.

\subsection{Predictive uncertainty quantification for each action}
For any subset $\mathcal{S}$, we provide uncertainty quantification for the optimal linear coefficients $\bm\delta_\mathcal{S}$ using the predictive distribution under $\mathcal{M}$. Specifically, we modify \eqref{action} to remove the expectation under $p_{\mathcal{M}}\{\bm{\tilde y}(\bm{\tilde X}, \bm{\tilde Z}) \mid \bm y\} $ and therefore preserve the predictive uncertainty quantification:  
\begin{equation}\label{uq-solve}
\bm{\tilde \delta}_\mathcal{S}  \coloneqq \arg\min_{\bm\delta_\mathcal{S}}
\mathbb{E}_{[\bm \psi \mid \bm y]} \mathcal{L}\{\bm{\tilde y}(\bm{\tilde X}, \bm{\tilde Z}), \bm{\delta}_\mathcal{S}; \bm\psi\}   
= (\bm {\tilde X}_\mathcal{S}'\bm{\hat \Omega} \bm{\tilde X}_\mathcal{S})^{-1} \bm{\tilde X}_\mathcal{S}'  \bm{\hat \Omega}\bm{\tilde y}(\bm{\tilde X}, \bm{\tilde Z}). 
\end{equation}
This mechanism for uncertainty quantification generalizes the predictive projection approach from \cite{Kowal2021} to account for Mahalanobis loss. In particular, \eqref{uq-solve} includes marginalization over $\bm{\Omega_\psi}$ to ensure that the resulting quantity is exclusively a \emph{posterior predictive} variable with a distribution induced by $p_{\mathcal{M}}\{\bm{\tilde y}(\bm{\tilde X}, \bm{\tilde Z}) \mid \bm y\} $. However, \eqref{uq-solve} can be modified to include the uncertainty of $\bm{\Omega_\psi}$ by replacing $\bm{\hat \Omega}$ with $\bm{\Omega_\psi}$. Posterior samples of $\bm{\tilde \delta}_\mathcal{S}$ only require posterior predictive samples of $\bm{\tilde y}(\bm{\tilde X}, \bm{\tilde Z})$---which can be shared among all subsets $\mathcal{S}$ of interest---and the solution to a GLS problem \eqref{uq-solve}. In particular, we use \eqref{uq-solve} to compute interval estimates for the linear coefficients associated with $\mathcal{S}_{min}$ and $\mathcal{S}_{small}$.

\section{Simulation study}\label{sims}
We evaluate the proposed LMM subset selection techniques using simulated data from a Gaussian random intercept model. First, we generate $p$ correlated fixed effects covariates from marginal standard normal distributions with $\mbox{Cor}(x_{i,j}, x_{i,j'}) = (0.75)^{\vert j - j' \vert}$ for $i=1,\ldots, n$ and $j=1,\ldots,p$. The $p$ columns are randomly permuted and augmented with an intercept. The true linear coefficients $\bm \beta^*$ are constructed by setting $\beta_0^* = -1$ and fixing $p_* = 5$ nonzero coefficients, with $\lceil p_*/2 \rceil$ equal to $1$ and $\lfloor p_*/2 \rfloor$ equal to $-1$, and the rest at zero. Let $y_i^* \coloneqq \bm x_i' \bm\beta^*$ denote the true expectation. For a given intraclass correlation $\rho_*$ and signal-to-noise ratio  SNR, define $\sigma_{tot}^2 \coloneqq \mbox{var}(\{y_i^*\}_{i=1}^n)/\mbox{SNR}$ and let $\sigma_u^2 \coloneqq {\rho_*\sigma_{tot}^2}$ and $\sigma_\epsilon^2 \coloneqq  {\sigma_{tot}^2 - \sigma_u^2}$. The data are generated as $y_{ij} = y_i^* + u_i + \epsilon_{ij}$ where $u_i \stackrel{iid}{\sim} N(0, \sigma_u^2)$ and $\epsilon_{ij} \sim N(0, \sigma_\epsilon^2)$ for $j=1,\ldots, m$ and $i=1,\ldots, n$.  We consider $\rho_* = 0.25$, $m = 4$, $\mbox{SNR} = 1$ (see the supporting information for SNR = 5), $p \in \{15, 200\}$, and $n \in \{75, 150, 300\}$.
 We repeat the data-generating process 100 times for each design.

We implement a Bayesian LMM using the sampler from Section~\ref{rand-int} and horseshoe priors on the fixed effects (see Section~\ref{app}). Using $\mathcal{M}$, we extract the acceptable family $\mathbb{A}_{0, 0.10}$ with $s_k=15$. We compute point predictions for all acceptable subsets and evaluate $\mathcal{S}_{min}$ and $\mathcal{S}_{small}$ for variable selection and uncertainty quantification via the 90\% intervals from $\bm{\tilde \delta}_\mathcal{S}$. 
The primary Bayesian competitor is given by the usual actions under  $\mathcal{M}$: posterior expectations for point predictions, 90\% highest posterior density (HPD) intervals of $\bm \beta$ for uncertainty quantification, and selection based on whether the 95\% HPD intervals for each $\beta_j$ exclude zero. As a secondary Bayesian competitor, we compute the point predictions and interval estimates from $\mathcal{S}_{small}$ under  a (non-LMM) Gaussian linear regression model using squared error loss \citep{Kowal2021}, which ignores the longitudinal aspect of the data. Lastly, we compare against classical selection methods that do not account for the random effects. Specifically, we apply the adaptive lasso (tuning parameter selected via 10-fold cross-validation and the one-standard-error rule) and classical subset selection (using AIC) to the data $\{(\bm x_i, \bar y_i)\}_{i=1}^n$ for $\bar y_i \coloneqq m^{-1} \sum_{j=1}^{m} y_{ij}$. We attempted to include fence-based variable selection for LMMs \citep{Jiang2008}, but the \texttt{R} package \texttt{fence} failed to converge for any simulation settings with $p \ge 10$.

Point prediction accuracy is evaluated using Mahalanobis loss for $y_i^*$, where the weight matrix \eqref{prec-int} uses the true parameters for $\sigma_\epsilon^2$ and $\sigma_u^2$. The simulation-averaged results are in 
Table~\ref{tab:accept-maha}. $\mathcal{S}_{small}$ consistently provides the best or near-best point predictions, followed by the posterior mean under $\mathcal{M}$. Further, $\mathcal{S}_{small}$  usually selects fewer variables than  all competitors (not shown).  Hence, $\mathcal{S}_{small}$ offers substantial reductions in the subset size while maintaining near-optimal prediction accuracy---which is precisely the goal of the smallest acceptable subset.

\begin{table}[ht] \small
\centering
\begin{tabular}{r|rrrrrr|rrrrr} 
  \hline
$(n, p)$ & lasso & subset & $\mathcal{M}$ & SE & $\mathcal{S}_{min}$ & $\mathcal{S}_{small}$ & $\mathbb{A}^{0}$ & $\mathbb{A}^{0.1}$ & $\mathbb{A}^{0.5}$ & $\mathbb{A}^{0.9}$ & $\mathbb{A}^{1}$ \\ 
  \hline
  $(75, 15)$  & 0.182 & 0.116 & 0.108 & 0.100 & 0.109 & {\bf 0.097} & 0.070 & 0.099 & 0.109 & 0.117 & 0.170 \\ 
  $(75, 200)$  & 0.507 & 0.644 & 0.426 & 0.351 & 0.410 & {\bf 0.315} & 0.279 & 0.375 & 0.399 & 0.410 & 0.444 \\ 
    $(150, 15)$  & 0.104 & 0.056 & 0.050 & 0.042 & 0.053 & {\bf 0.037} & 0.032 & 0.048 & 0.054 & 0.058 & 0.066 \\ 
  $(150, 200)$  & 0.275 & 0.341 & 0.127 & 0.151 & 0.194 & {\bf 0.111} & 0.108 & 0.160 & 0.184 & 0.196 & 0.201 \\ 
    $(300, 15)$  & 0.057 & 0.025 & 0.023 & 0.019 & 0.025 & {\bf 0.015} & 0.014 & 0.022 & 0.026 & 0.027 & 0.028 \\  
  $(300, 200)$  & 0.096 & 0.203 & {\bf 0.048} & 0.078 & 0.105 & 0.056 & 0.054 & 0.083 & 0.099 & 0.107 & 0.110 \\ 
    \hline 
\end{tabular}
\caption{ \small
{\bf Left:} average Mahalanobis loss for each method; the smallest (best)  value is bolded. {\bf Right:} for each simulation, we  compute the $q$th quantile of the true Mahalanobis loss for predicting $y_i^*$ for all acceptable subsets, and then average that quantity across simulations to obtain $\mathbb{A}^{q}$. $\mathcal{M}$ refers to the posterior mean under the Bayesian LMM and SE  is the smallest acceptable subset using linear regression and squared error loss \citep{Kowal2021}. $\mathcal{S}_{small}$ performs exceptionally well across all settings, while  even the worst acceptable subsets $(\mathbb{A}^1)$ outperform the frequentist competitors.
\label{tab:accept-maha}}
\end{table}

More broadly, we evaluate the overall predictive performance of the acceptable family $\mathbb{A}_{0, 0.10}$ using  the $q$th quantile of the true Mahalanobis loss for each acceptable subset at each simulation, and then average that quantity across simulations to obtain $\mathbb{A}^{q}$. For example, $\mathbb{A}^{1}$ is the worst possible performance in the acceptable family, i.e., if an oracle were to select the \emph{worst acceptable} subset at each simulation. The main takeaways (see Table~\ref{tab:accept-maha}) are (i) $\mathcal{S}_{small}$ typically outperforms even $\mathbb{A}^{0.1}$ and therefore is consistently in the top 10\% of acceptable subsets, and (ii) even the worst acceptable subsets  outperform the frequentist competitors. These results confirm our notion of \emph{near-optimality} of the acceptable family.

\begin{figure}[h!]
\begin{center}
\includegraphics[width=.32\textwidth]{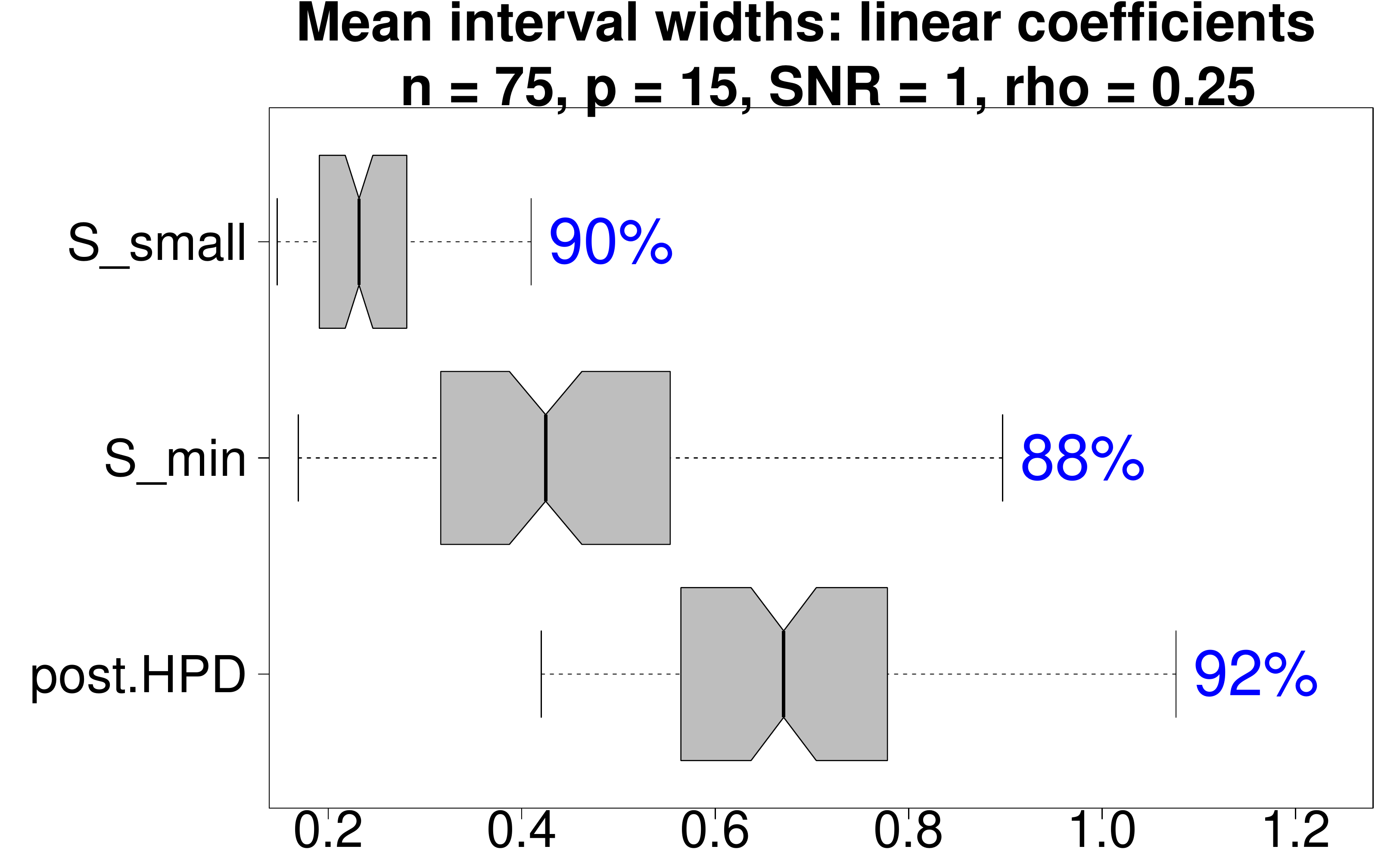}
\includegraphics[width=.32\textwidth]{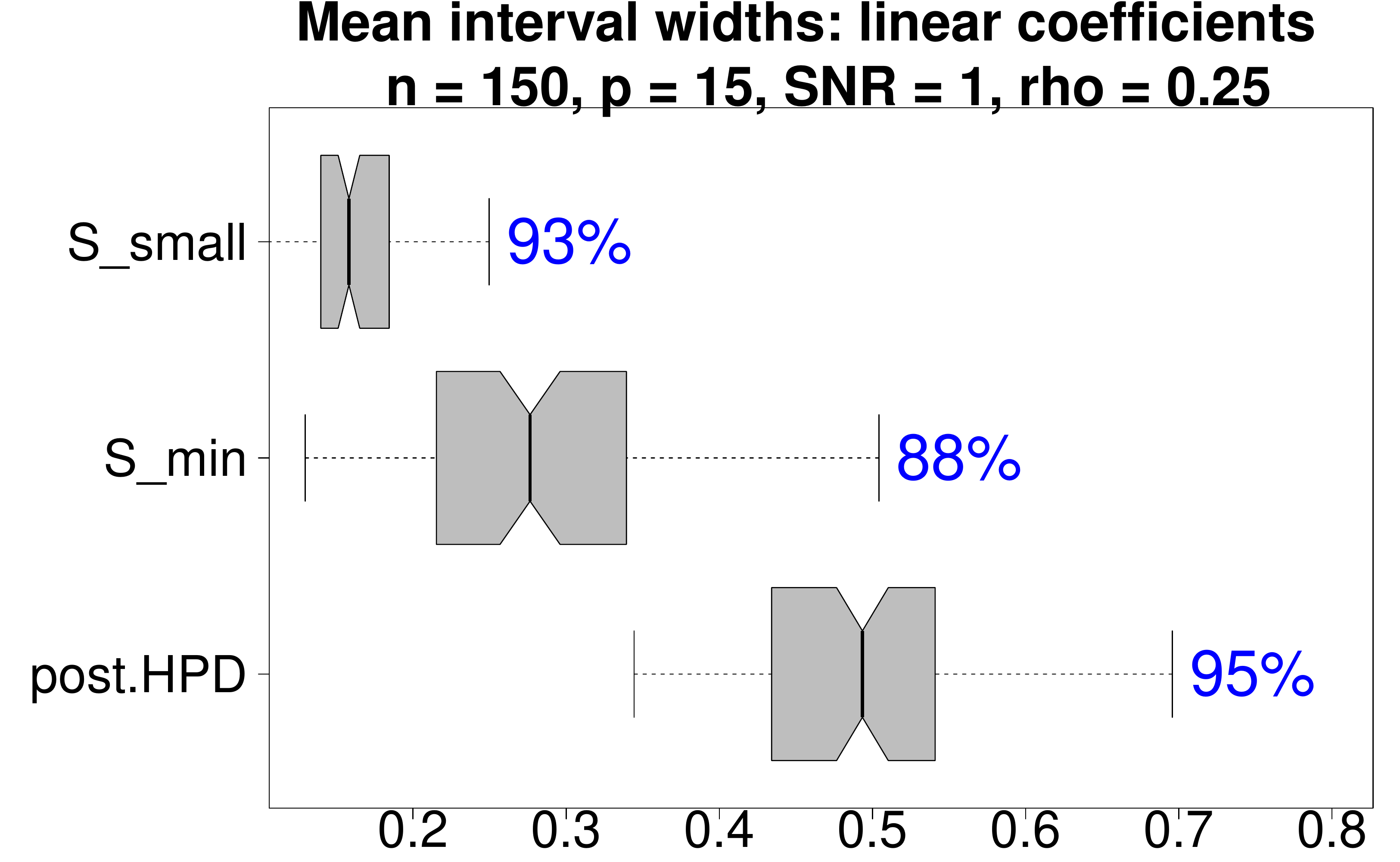}
\includegraphics[width=.32\textwidth]{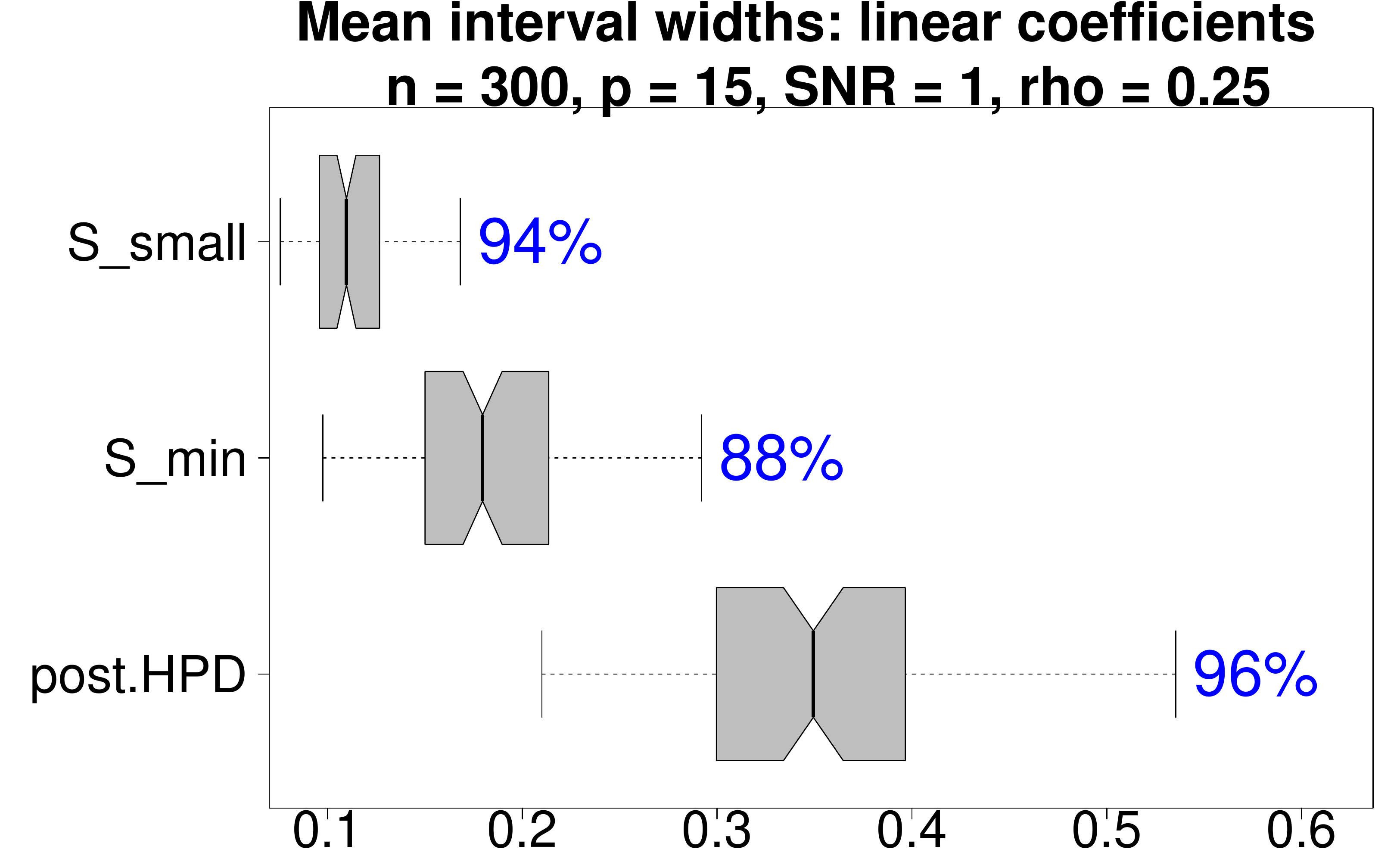}

\includegraphics[width=.32\textwidth]{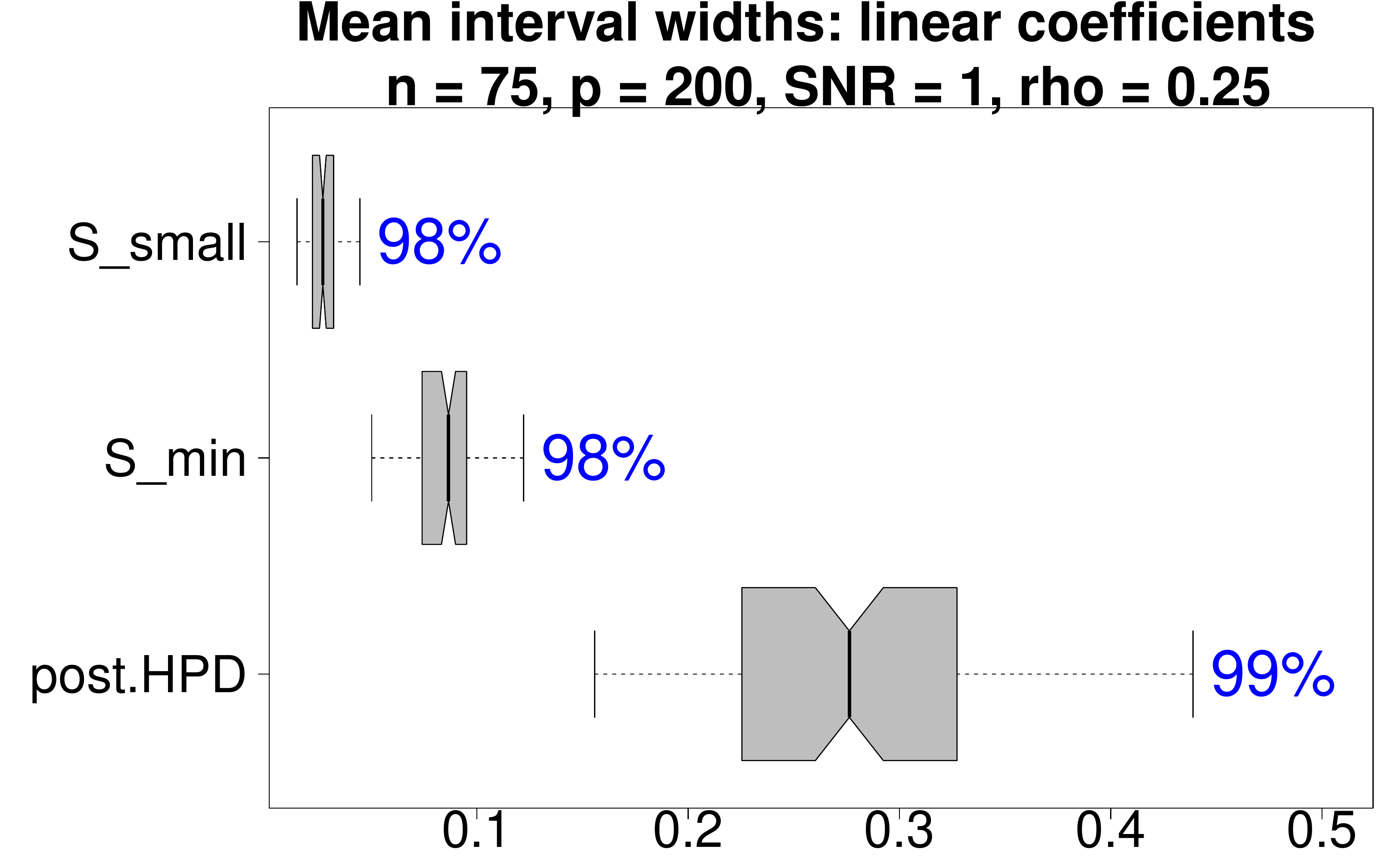}
\includegraphics[width=.32\textwidth]{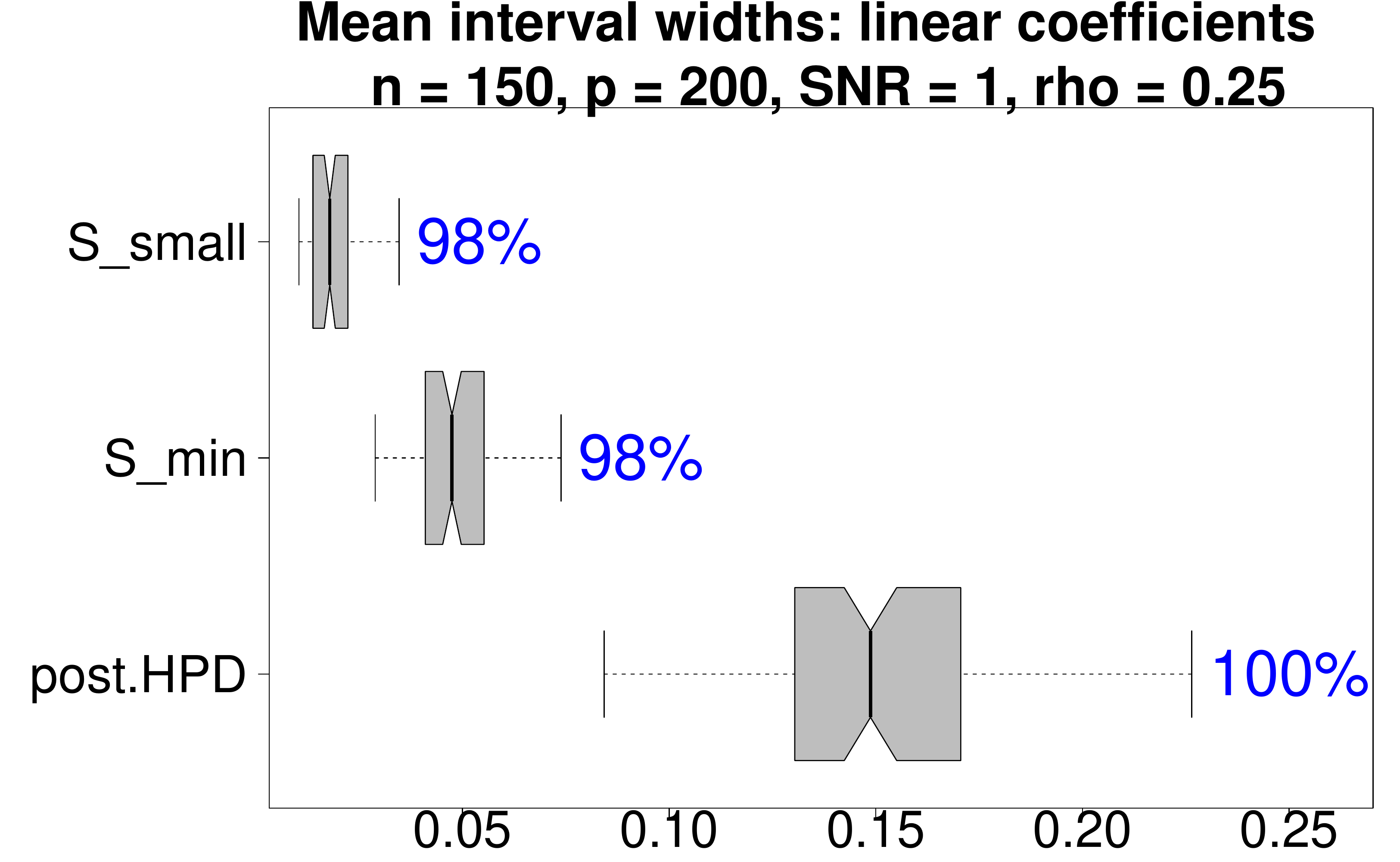}
\includegraphics[width=.32\textwidth]{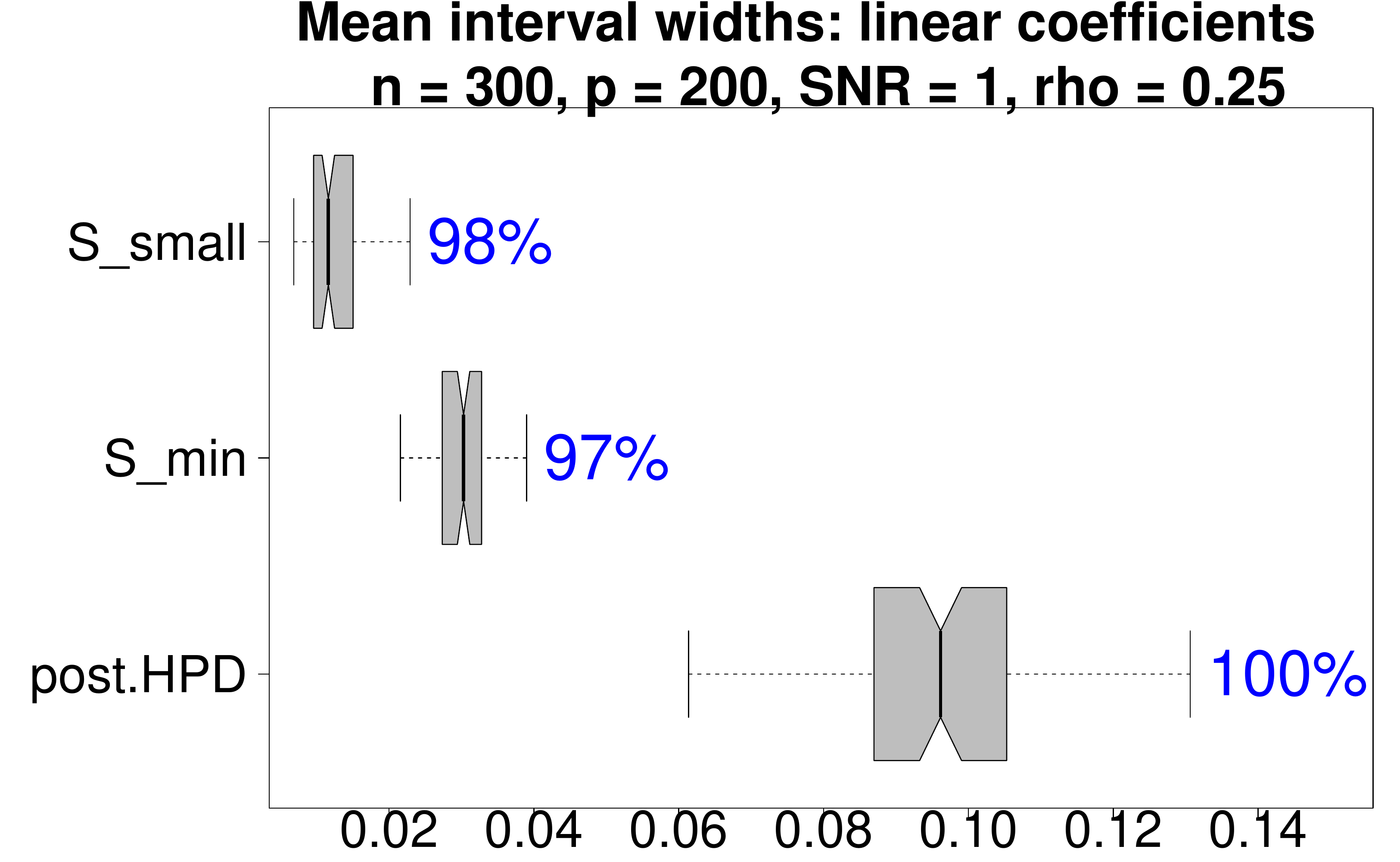}
\end{center}
\caption{ \small 
Mean 90\% interval widths (boxplots) with empirical coverage (annotations) for $\bm \beta^*$. Non-overlapping notches indicate significant differences between medians. The proposed intervals based on $\mathcal{S}_{small}$ are significantly narrower than the usual HPD intervals under $\mathcal{M}$ yet maintain the nominal 90\% coverage. 
\label{fig:sims-int}}
\end{figure}

The 90\% interval estimates for $\bm\beta^*$ are evaluated in Figure~\ref{fig:sims-int}, which reports the mean interval widths and the empirical coverage; narrow intervals that provide the correct nominal coverage are preferred. The intervals from $\mathcal{S}_{small}$ are clearly the best for these cases: the intervals maintain 90\% coverage and are much narrower than competing methods. In particular, the intervals from both  $\mathcal{S}_{small}$ and $\mathcal{S}_{min}$  are far more precise (i.e., narrower) than the 90\% HPD intervals under $\mathcal{M}$. 

Lastly, we evaluate the (marginal)  selection capabilities using true positive rates (TPRs) and true negative rates (TNRs) in Table~\ref{tab:tprs}.   $\mathcal{S}_{small}$ provides consistently high TPRs and TNRs, while the 95\% HPD intervals under $\mathcal{M}$ are far too conservative for selection (low TPRs). Both selection mechanisms are based on the same Bayesian LMM $\mathcal{M}$, but  $\mathcal{S}_{small}$ is decisively better. The improvements over classical subset selection are also substantial.

\begin{table}[ht] \small
\centering
\begin{tabular}{rrrrrrr}
  \hline
$(n, p) $ & & lasso & subset & posterior HPD & $\mathcal{S}_{min}$ & $\mathcal{S}_{small}$ \\ 
  \hline
  \multirow{2}*{ $(75, 15)$ } & TPR  & 0.95 & 0.98 & 0.86 & 0.99 & 0.95 \\ 
  & TNR  & 0.93 & 0.80 & 0.98 & 0.63 & 0.95 \\ 
      \hline 
      \multirow{2}*{ $(75, 200)$ } & TPR  & 0.94 & 0.91 & 0.57 & 0.94 & 0.91 \\
  & TNR  & 0.96 & 0.92 & 1.00 & 0.91 & 0.98 \\ 
        \hline 
    \multirow{2}*{ $(150, 15)$ } & TPR  & 0.99 & 1.00 & 0.99 & 1.00 & 0.99 \\ 
  & TNR  & 0.98 & 0.82 & 0.99 & 0.68 & 0.97 \\ 
    \hline 
      \multirow{2}*{ $(150, 200)$ } &  TPR & 0.99 & 1.00 & 0.95 & 1.00 & 1.00 \\ 
  & TNR  & 0.97 & 0.93 & 1.00 & 0.92 & 0.98 \\ 
        \hline 
    \multirow{2}*{ $(300, 15)$ } & TPR  & 1.00 & 1.00 & 1.00 & 1.00 & 1.00 \\ 
  & TNR  & 1.00 & 0.85 & 0.99 & 0.69 & 0.98 \\
    \hline 
      \multirow{2}*{ $(300, 200)$ } & TPR  & 1.00 & 1.00 & 1.00 & 1.00 & 1.00 \\
  & TNR  & 0.99 & 0.92 & 1.00 & 0.92 & 0.98 \\ 
    \hline 
\end{tabular}
\caption{ \small
True positive rates (TPR) and true negative rates (TNR) for synthetic data with $p_* + 1= 6$ active covariates including the intercept. $\mathcal{S}_{small}$ provides consistently high TPRs and TNRs, while the 95\% HPD intervals under $\mathcal{M}$ are far too conservative (low TPRs for smaller $n$). 
\label{tab:tprs}}
\end{table}

\section{Application}\label{app}
We apply our subset selection analysis to moderate-to-vigorous physical activity (MVPA) data from NHANES 2005-2006. Intraday activity was measured on each subject using hip-worn accelerometers for one to seven days. $\mbox{MVPA}_{ij}$   is defined as the number of minutes with at least 2020 activity counts for subject $i$ on day $j$, and typically corresponds to more intensive activities that include vigorous walking or running  \citep{Fishman2016}. The goal is to determine the subject-specific factors that predict MVPA. However, these longitudinal data feature repeated measurements on each participant, and this within-subject dependence must be accounted for in both modeling and decision analysis.

We specifically analyze older (ages 65-80) and Hispanic (Mexican American or Other Hispanic) individuals. Fixed effects include body mass index (BMI), age, gender (male or female), education level (less than high school, completed high school only, or some college and above), total cholesterol, HDL cholesterol, systolic blood pressure, smoking status (never, former, or current), drinking status (never, moderate, or heavy), and presence of diabetes.  After filtering to individuals with at least one day of activity data, days with at least 10 hours of accelerometer wear time, activity measurements that were correctly ``calibrated" and ``reliable" as flagged by NHANES, and individuals with no mobility problems, the resulting analysis dataset has $N = 243$ measurements on   $n = 61$ individuals with $p=13$ covariates.

We model $y_{ij} = \log(\mbox{MVPA}_{ij} + 1)$ using a Gaussian random intercept model (see Section~\ref{rand-int}) with horseshoe priors for the fixed effects $\bm \beta$, a Jeffreys prior for $\sigma_\epsilon^2$, and a uniform prior for $\sigma_u \sim \mbox{Unif}(0, 100)$. The Gibbs sampler generated 10,000 samples after a burn-in of 5,000; traceplots indicated no lack of convergence and the effective sample sizes were sufficiently large. A 95\% HPD interval for the within-subject correlation, $\sigma_u^2/(\sigma_u^2 + \sigma_\epsilon^2)$, is $(0.18, 0.48)$, which suggests moderate within-subject autocorrelation.

\begin{figure}[h!]
\begin{center}
\includegraphics[width=.8\textwidth]{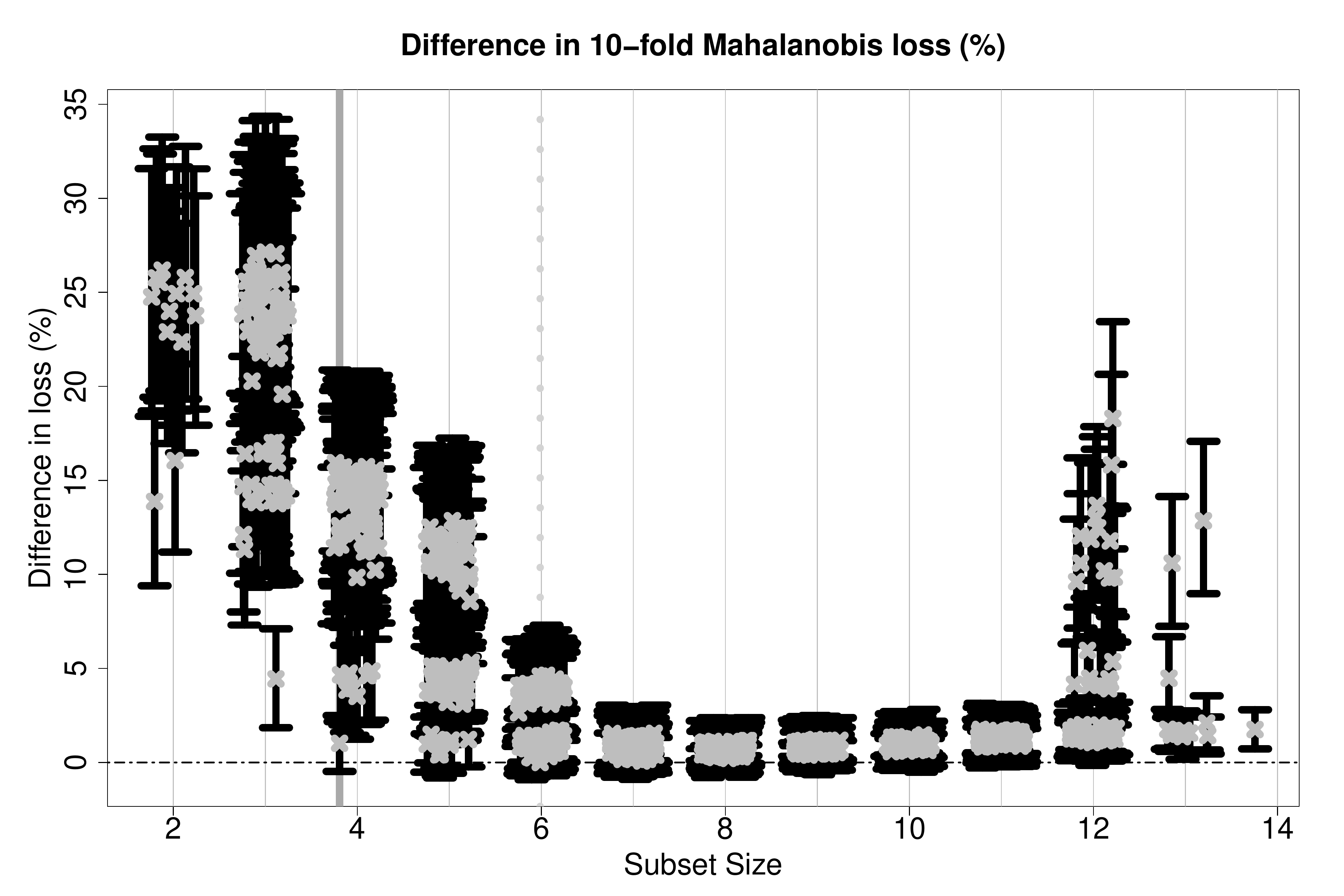}
\end{center}
\caption{ \small 
Prediction 80\% intervals (lines) and expectations (points) for  $\widetilde{\mathcal{D}}_{\mathcal{S}_{min},\mathcal{S}}$ across candidate subsets  $\mathcal{S}$ (including the intercept). The analogous empirical quantity, ${\mathcal{D}}_{\mathcal{S}_{min},\mathcal{S}}  \coloneqq 100\times({\mathcal{L}}_{\mathcal{S}} - {\mathcal{L}}_{\mathcal{S}_{min}})/{\mathcal{L}}_{\mathcal{S}_{min}}$, is also denoted (x-marks). Subsets of the same size are jittered for clarity of presentation. Intervals that include $\eta = 0$ (horizontal line) correspond to acceptable subsets. Acceptable subsets range from sizes 4 to 12, including $\vert \mathcal{S}_{small}\vert = 4$ (solid line) and $\vert  \mathcal{S}_{min}\vert = 6$ (dashed line). 
\label{fig:pred-loss}}
\end{figure}

Using the posterior and predictive samples from the Bayesian LMM, we compute and study the acceptable family. Since $p$ is not large, we filter from the $2^p  = 4096$ possible subsets to the ``best" $s_k=100$ models of each size $k=1,\ldots,s_{max} = p+1$ (the intercept  is always included), which produces 973 candidate subsets. Figure~\ref{fig:pred-loss} summarizes the predictive performance among these candidates using Mahalanobis predictive loss. The 80\% intervals that include $\eta = 0$ (horizontal line) correspond to acceptable subsets with $\varepsilon = 0.1$. Each subset of size two performs 10-35\% worse than $\mathcal{S}_{min}$. One subset of size three outperforms the rest and is within 3-7\% of $\mathcal{S}_{min}$; this subset  would be acceptable for margins $\eta > 3\%$ or smaller $\varepsilon \le 0.01$, which corresponds to wider  intervals in Figure~\ref{fig:pred-loss}.  The smallest acceptable subset for $\eta = 0\%$ has four variables, and notably performs  as well or better than the larger subsets; this subset is also unchanged for $\varepsilon \in [0.02, 0.16]$.

\begin{figure}[h!]
\begin{center}
\includegraphics[width=.8\textwidth]{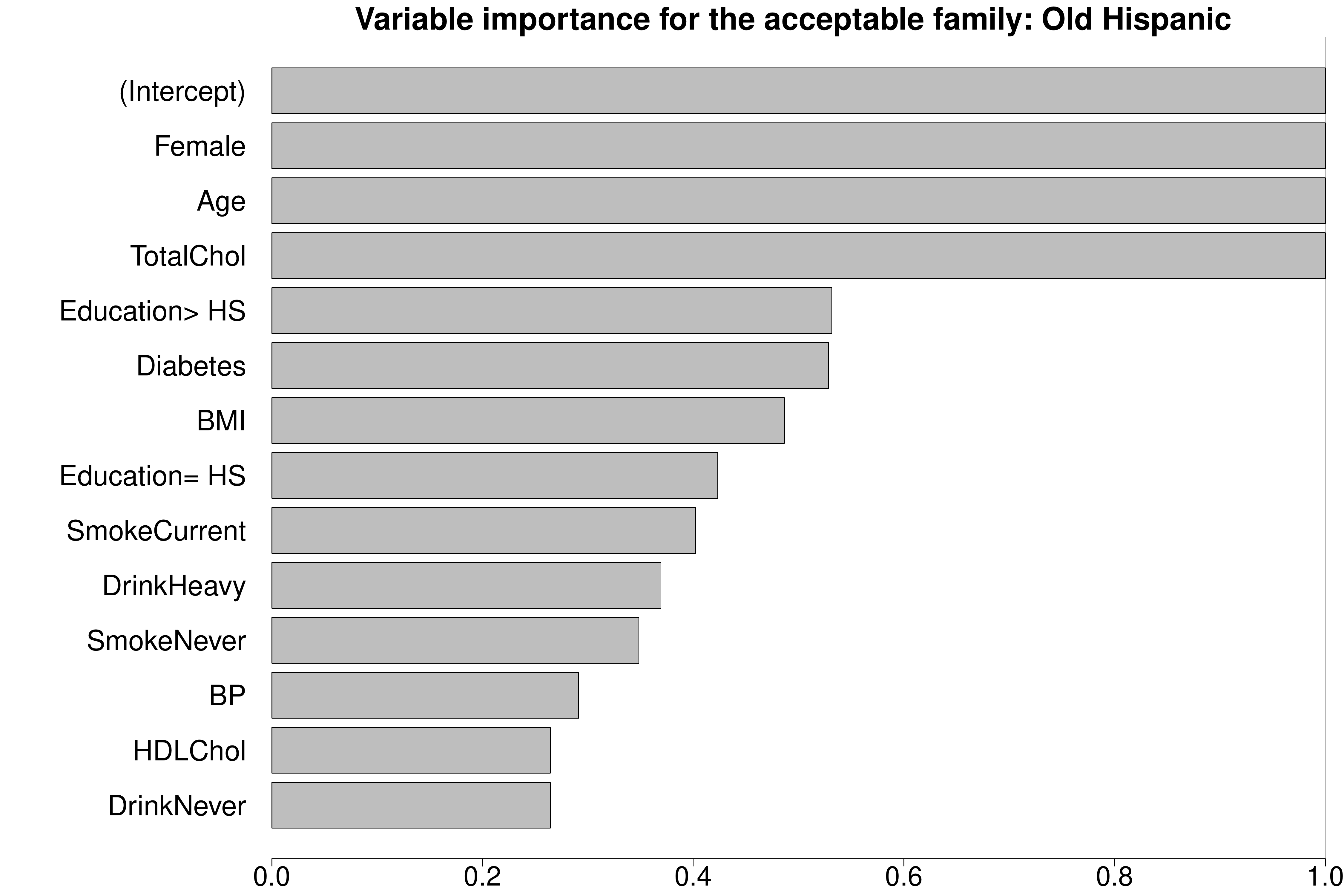}
\end{center}
\caption{ \small 
Variable importance $\mbox{VI}_{\rm incl}(j)$ for the acceptable family $\mathbb{A}_{0, 0.10}$. Gender, age, and total cholesterol are keystone covariates, while the other covariates appear in some---but not most---of the acceptable subsets.   These results are robust for $\varepsilon \in [0.02, 0.16]$.
\label{fig:vi}}
\end{figure}

The acceptable family features  $\vert \mathbb{A}_{0, 0.10}\vert =333$ members ranging from sizes 4 to 12. To summarize $\mathbb{A}_{0, 0.10} $, we report the variable importance metric $\mbox{VI}_{\rm incl}(j)$ in Figure~\ref{fig:vi}. Gender, age, and total cholesterol are keystone covariates that appear in all acceptable subsets, and are the only members (plus the intercept) of $\mathcal{S}_{small}$. Notably, the remaining covariates appear in some---but not most---of the acceptable subsets. These covariates are not entirely extraneous, but appear to be interchangeable and not strictly necessary for acceptable linear prediction. The ``best" subset $\mathcal{S}_{min}$ adds smoking status (current) and diabetes  to $\mathcal{S}_{small}$. Yet the variable importance provides important context for $\mathcal{S}_{min}$: although smoking status belongs to the ``best" subset, it only appears in a moderate fraction (about 40\%) of the acceptable subsets. By comparison, education level (some college and above) appears in vastly more acceptable subsets, yet does not belong to $\mathcal{S}_{small}$.

\begin{figure}[h!]
\begin{center}
\includegraphics[width=.8\textwidth]{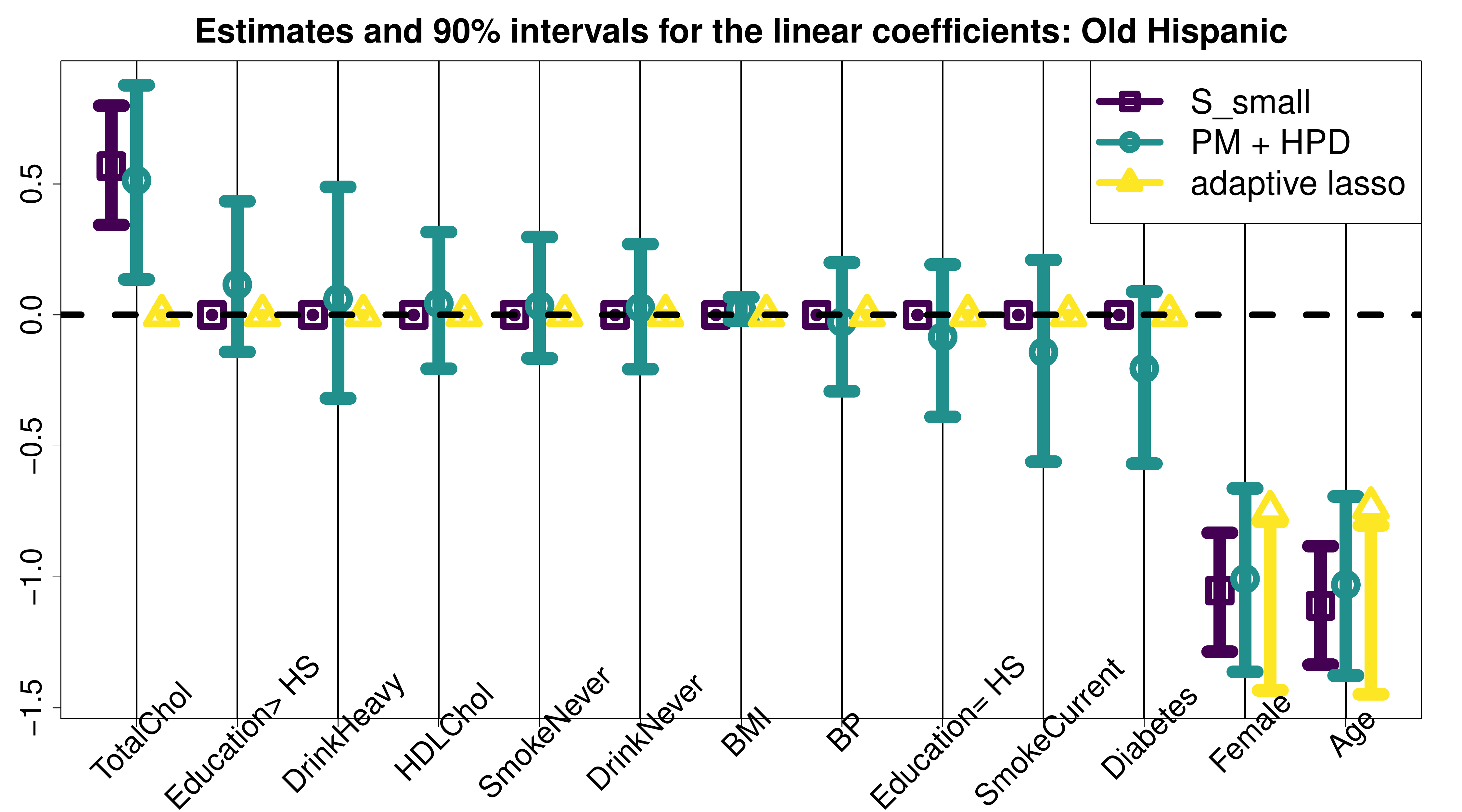}
\end{center}
\caption{ \small 
Estimated coefficients and 90\%  intervals for $\mathcal{S}_{small}$ ($\bm{\hat \delta}_{\mathcal{S}_{small}}$ and quantiles from $\bm{\tilde \delta}_{\mathcal{S}_{small}}$), the Bayesian LMM $\mathcal{M}$ (posterior means and HPD intervals for $\bm \beta$), and the adaptive lasso (intervals from \citealp{Zhao2021}). There is broad agreement, although $\mathcal{S}_{small}$ produces narrower intervals and sparse estimates compared to the the usual Bayesian LMM estimates.  These results are robust for $\varepsilon \in [0.02, 0.16]$.
\label{fig:coef}}
\end{figure}

Lastly, Figure~\ref{fig:coef} compares the point and interval estimates from $\mathcal{S}_{small}$ against the Bayesian LMM  $\mathcal{M}$ and the adaptive lasso. Both $\mathcal{S}_{small}$ and $\mathcal{M}$ highlight a positive effect for total cholesterol---perhaps a realization of the common advice that individuals with high cholesterol should exercise more---while all three methods agree on negative effects for gender (female) and age. $\mathcal{S}_{small}$ produces narrower intervals among the nonzero coefficients compared to the HPD intervals under $\mathcal{M}$, and offers a sparsity in point estimation that is not available for the posterior means under $\mathcal{M}$. Yet 
the  methods broadly agree: the selected variables in $\mathcal{S}_{small}$ correspond exactly to the 90\% HPD intervals under $\mathcal{M}$ that exclude zero. 
The frequentist  intervals from \cite{Zhao2021} are difficult to interpret, since they often fail to include the (adaptive) lasso-based point estimates from which they were derived.

The results are robust to $\varepsilon$: $\mathcal{S}_{small}$ is unchanged for $\varepsilon \in [0.02, 0.16]$, and the variable importances are stable. $\mathcal{S}_{small}$ omits total cholesterol for $\varepsilon = 0.01$ and adds  diabetes for $\varepsilon = 0.20$. The number of acceptable subsets decreases from $\vert \mathbb{A}_{0, \epsilon}\vert \in \{658, 456, 333,  60\} $ for $\varepsilon \in \{0.01, 0.05, 0.10, 0.20\}$, which is expected: larger values of $\varepsilon$ provide more lenient admission to the acceptable family.

Note that NHANES data are collected from a complex sampling design, and population-level inference typically requires survey adjustments. The oversampled groups in NHANES 2005-2006 are specific age groups (12-19 and 60+ years), races (Black and Mexican Americans), and low-income individuals. Because we subset by age group and race and further include age and many other covariates in the model, we expect that the effects of the sampling design are mitigated.

\section{Discussion}\label{concl}
We have developed a decision analysis strategy for subset selection in Bayesian LMMs. Using a Mahalanobis predictive loss function to bring forward the structured dependence from the LMM into the decision analysis, we derived optimal linear coefficients for (i) any \emph{given} subset of variables and (ii) all subsets of variables that satisfy a \emph{cardinality constraint}. The  coefficients  are accompanied by predictive uncertainty quantification and regularization inherited from the underlying Bayesian LMM. Comparing across subsets, we collected and summarized the \emph{acceptable family} of subsets that (nearly) matched the predictive performance of the ``best" subset. The proposed tools demonstrated excellent prediction, estimation, and selection properties on simulated data, and were applied to a longitudinal dataset to study the key predictors of MVPA.

Given the acceptable family of near-optimal subsets, it is natural to ask: ``Which subset should be used?" Our response is that,  based on  predictive accuracy, \emph{any} of the acceptable subsets provides a reasonable answer. Absent additional information (such as  individual variable costs), we advocate the \emph{smallest} acceptable subset $\mathcal{S}_{small}$, which simultaneously (i) provides excellent prediction, uncertainty quantification, and selection capabilities across a variety of challenging simulation settings and (ii) 
offers a notion of the necessary variables for near-optimal linear  prediction (when $\mathcal{S}_{small}$ is unique), since smaller subsets are \emph{not} acceptable by definition. However, our prioritization of P4  underlines the crucial point that no \emph{single} subset---including $\mathcal{S}_{small}$---should be used in isolation to report the variables that ``matter". In particular, variables excluded from the ``best" subset are not necessarily  irrelevant, while variables included in  the ``best" subset  are  not necessarily essential. The acceptable family fills in those gaps to provide a more complete picture, and is accompanied by suitable summaries.

The Mahalanobis loss \eqref{loss} is designed for the LMM \eqref{lme}, which is most commonly a Gaussian LMM.  Although we focused primarily on random intercept and random slope models, the results are broadly applicable among LMMs, including many functional data and spatial regression models. In addition, modifications for non-Gaussian \emph{generalized}  LMMs (GLMMs) may be attainable. For Bayesian subset selection with binary data, \cite{Kowal2021} used iteratively-reweighted least squares (IRLS) to approximate the minimizer of a cross-entropy loss with a weighted least squares solution. IRLS is a widely popular strategy for estimating generalized linear models, and can be used to produce optimal coefficients under the corresponding deviance loss functions. For LMMs, a natural modification would be to insert a weighting matrix akin to $\bm{\Omega_\psi}$ into the IRLS, thereby extending the proposed tools for compatibility with GLMMs.

\subsection*{Acknowledgements}
Research was sponsored by the Army Research Office and was accomplished under Grant Number W911NF-20-1-0184. The views and conclusions contained in this document are those of the authors and should not be interpreted as representing the official policies, either expressed or implied, of the Army Research Office or the U.S. Government. The U.S. Government is authorized to reproduce and distribute reprints for Government purposes notwithstanding any copyright notation herein. 

\bibliographystyle{apalike}
\bibliography{refs.bib}

\end{document}